\documentclass[useAMS,usenatbib]{mn2e}

\pdfoutput=1
\usepackage{epsfig}
\usepackage{myaasmacros}
\usepackage{amsmath}
\usepackage{amssymb}
\usepackage{graphicx}
\usepackage{deluxetable}

\title[Origin of fast and slow rotators]
{The ATLAS$^{\rm{3D}}$ project - XXV: Two-dimensional kinematic
  analysis of simulated galaxies and the cosmological origin of fast
and slow rotators}
\author[T.~Naab et al.]
{Thorsten Naab\thanks{email: naab@mpa-garching.mpg.de}$^{1}$,
L. Oser$^{1,2}$,
E. Emsellem$^{3,4}$,
M. Cappellari$^{5}$,
D. Krajnovi\'c$^{6}$,\newauthor
R. M. McDermid $^{7}$,
K. Alatalo$^8$,
E. Bayet$^5$, 
L. Blitz$^8$,
M. Bois$^{7}$,
F. Bournaud$^{10}$,\newauthor
M. Bureau$^{5}$,
A. Crocker$^{11}$,
R. L. Davies$^{3}$,
T. A. Davis$^{3,4}$,
P. T. de Zeeuw$^{3,12}$,\newauthor
P.-A. Duc$^{10}$,
M. Hirschmann $^{13}$,
P. H. Johansson$^{14}$,
S. Khochfar$^{15}$,
H. Kuntschner$^{16}$,\newauthor
R. Morganti$^{17,18}$,
T. Oosterloo$^{17,18}$
M. Sarzi$^{19}$,
N. Scott$^{20}$,
P. Serra$^{17,24}$,
G. van de Ven$^{21}$,\newauthor
A. Weijmans \thanks{Dunalp Fellow}$^{22}$,
and L. M. Young$^{23}$\\
$^{1}$Max-Planck Institut f\"ur Astrophysik, Karl-Schwarzschild-Str. 1, 85741 Garching, Germany\\
$^{2}$ Department of Astronomy, Columbia University, Pupin Physics
Laboratories, New York, NY 10027, USA\\
$^{3}$European Southern Observatory, Karl-Schwarzschild-Str. 2, 85748 Garching, Germany\\
$^{4}$Universit\'e Lyon 1, Observatoire de Lyon, Centre de Recherche
Astrophysique de Lyon and Ecole Normale Sup\'erieure de Lyon, \\
9 avenue Charles Andr\'e, F-69230 Saint-Genis Laval, France\\ 
$^{5}$Sub-department of Astrophysics, University of Oxford, Denys Wilkinson Building, Keble Road, Oxford OX1 3RH\\
$^{6}$Leibnitz-Institut f\"ur Astrophysik Potsdam (AIP), An der Sternwarte 16, D-14482 Potsdam, Germany\\
$^{7}$Gemini Observatory, Northern Operations Centre, 670 N. A`ohoku Place, Hilo, HI 96720, USA\\
$^{8}$Department of Astronomy, Campbell Hall, University of California, Berkeley, CA 94720, USA\\
$^{9}$Laboratoire AIM Paris-Saclay, CEA/IRFU/SAp  CNRS  Universit\'e Paris Diderot, 91191 Gif-sur-Yvette Cedex, France\\
$^{10}$ Observatoire de Paris, LERMA and CNRS, 61 Av. de
l'Observatoire, F-75014 Paris, France\\
$^{11}$Department of Astronomy, University of Massachusetts, Amherst, MA 01003, USA)\\
$^{12}$Sterrewacht Leiden, Leiden University, Postbus 9513, 2300 RA
Leiden, the Netherlands\\
$^{13}$INAF Astronomical Observatory of Trieste, via G. B. Tiepolo 11, I-34143 Trieste, Italy \\
$^{14}$ Department of Physics, University of Helsinki, Gustaf H\"allstr\"omin katu 2a, FI-00014 Helsinki, Finland\\
$^{15}$Max-Planck Institut f\"ur extraterrestrische Physik, PO Box 1312, D-85478 Garching, Germany\\
$^{16}$Space Telescope European Coordinating Facility, European Southern Observatory, Karl-Schwarzschild-Str. 2, 85748 Garching, Germany\\
$^{17}$Netherlands Institute for Radio Astronomy (ASTRON), Postbus 2, 7990 AA Dwingeloo, The Netherlands\\
$^{18}$Kapteyn Astronomical Institute, University of Groningen, Postbus 800, 9700 AV Groningen, The Netherlands\\
$^{19}$Centre for Astrophysics Research, University of Hertfordshire, Hatfield, Herts AL1 9AB, UK\\
$^{20}$Centre for Astrophysics and Supercomputing, Swinburne University of Technology, PO Box 218, Hawthorn, VIC 3122, Australia \\
$^{21}$Max-Planck Institut f\"ur Astronomie, K\"onigstuhl 17, D-69117 Heidelberg, Germany\\
$^{22}$School of Physics and Astronomy, University of St Andrews,
North Haugh, St Andrews, Fife KY16 9SS\\
$^{23}$Physics Department, New Mexico Institute of Mining and Technology, Socorro, NM 87801, USA \\
$^{24}$CSIRO Astronomy and Space Science, Australia Telescope National Facility, PO Box 76, Epping, NSW 1710, Australia
}
\date{Accepted ???. Received ??? in original form ???}

\pagerange{\pageref{firstpage}--\pageref{lastpage}} \pubyear{2012}

\begin{document}
\label{firstpage}
\maketitle
\clearpage 

\begin{abstract} 
We present a detailed two-dimensional stellar dynamical
analysis of a sample of 44 cosmological hydrodynamical simulations of
individual central galaxies with stellar masses of $2 \times 10^{10} M_{\odot}
\lesssim M_* \lesssim 6 \times 10^{11} M_{\odot}$. Kinematic maps of
the stellar line-of-sight velocity, velocity dispersion, and
higher-order Gauss-Hermite moments $h_3$ and $h_4$ are constructed for
each central galaxy and for the most massive satellites. The amount of
rotation is quantified using the $\lambda_{\mathrm{R}}$-parameter. The
velocity, velocity dispersion, $h_3$, and $h_4$ fields of the
simulated galaxies show a diversity similar to observed kinematic maps
of early-type galaxies in the ATLAS$^{\rm{3D}}$ survey. This includes fast
(regular), slow, and misaligned rotation, hot spheroids with embedded
cold disk components as well as galaxies with counter-rotating cores
or central depressions in the velocity dispersion. We link the present
day kinematic properties to the individual cosmological formation
histories of the  galaxies. In general, major galaxy mergers have a 
significant influence on the rotation properties resulting in
both a spin-down as well as a spin-up of the merger remnant. Lower mass
galaxies with significant ($\gtrsim$ 18 per cent) in-situ formation of stars
since $z \approx 2$, or with additional gas-rich major mergers -
resulting in a spin-up - in their formation history, form elongated
($\epsilon \sim 0.45$) fast rotators ($\lambda_{\mathrm{R}} \sim
0.46$) with a clear anti-correlation of  $h_3$ and  $v/\sigma$. An
additional formation path for fast rotators includes gas poor major
mergers leading to a spin-up of the remnants ($\lambda_{\mathrm{R}}
\sim 0.43$). This formation path does not result in anti-correlated
$h_3$ and $v/\sigma$. The formation histories of slow rotators can
include late major mergers. If the merger is gas-rich the remnant
typically is a less flattened slow rotator with a central dip in the
velocity dispersion. If the merger is gas poor the remnant is very
elongated ($\epsilon \sim 0.43$) and slowly rotating
($\lambda_{\mathrm{R}} \sim 0.11$). The galaxies most consistent with
the rare class of non-rotating round early-type galaxies grow by gas-poor
minor mergers alone. In general, more massive galaxies have less
in-situ star formation since $z \sim 2$, rotate slower and have older
stellar populations. We discuss general implications for the formation
of fast and slowly rotating galaxies as well as the weaknesses and
strengths of the underlying models.  
\end{abstract}

\begin{keywords}
ISM: clouds --- ISM: kinematics and dynamics --- stars: formation
\end{keywords}

\section{Introduction}
\label{intro}
Observationally, the $ATLAS^{\rm{3D}}$ survey
\citep{2011MNRAS.413..813C} provides the most complete panoramic view 
on the properties of 260 local early-type galaxies in a volume
limited sample covering different environments within a distance of
$\sim 42 \rm Mpc$. This includes a complete inventory of the 
central \citep{2011MNRAS.414.2923K,2011MNRAS.414..888E} and extended
\citep{2011MNRAS.417..863D,2013MNRAS.432.1796A} baryonic galactic building blocks such as stars,
molecular gas \citep{2011MNRAS.414..940Y,2011MNRAS.414..968D}, neutral
gas \citep{2012MNRAS.422.1835S} and ionized gas
\citep{2011MNRAS.417..882D}, as well as high-density gas tracers
\citep{2012MNRAS.421.1298C}. This is combined with unique
two-dimensional information  about the stellar
\citep{2011MNRAS.414.2923K,2011MNRAS.414..888E} and gaseous 
\citep{2011MNRAS.414..968D} kinematics as well as photometry
\citep{2011MNRAS.414.2923K,2013MNRAS.432.1768K,2013arXiv1305.4973K}
within the main body of the galaxies.     

The theoretical effort within the survey is twofold. Based on the
observed photometry and kinematics we aim at understanding the
underlying three-dimensional dynamical structure
(\citealp{2012MNRAS.424.1495L,2012Natur.484..485C,2013MNRAS.432.1709C,2013MNRAS.432.1862C})
as well as the spatially resolved chemical 
composition and ages of the stellar populations (\citealp{2013MNRAS.432.1894S}; McDermid et al. 2013, in
preparation; Kuntschner et
al. 2013, in preparation). With a - backwards - archaeological
approach we can then put tentative constraints on the formation
histories of early-type galaxies. Using semi-analytical models
\citep{2011MNRAS.417..845K} and direct numerical simulations we
also investigate possible - forward - formation scenarios and check their
success in predicting the observed present day galaxy properties. The
simulations cover different levels of complexity: idealized
high-resolution simulations of mergers between two or more galaxies
including (or not) star formation
\citep{2010MNRAS.406.2405B,2011MNRAS.416.1654B}, 
simulations of model realisations of observed galaxies
\citep{2012arXiv1212.2288M}, and simulations of the entire formation 
history of galaxies in a full cosmological context as presented here.

One of the striking results from $ATLAS^{\rm{3D}}$, which is also the focus
of this paper, is that only a small fraction (12\%, 32/260) of the
galaxies rotate slowly with no indication of embedded disc
components. In contrast, the majority (86\%, 224/260) of early-type galaxies shows
significant (disc-like) rotation with regular velocity fields
\citep{2011MNRAS.414.2923K,2011MNRAS.414..888E}.  Galaxies with 
corresponding properties were coined slow rotators and fast rotators,
respectively, by the preceding SAURON survey
\citep{2001MNRAS.326...23B,2002MNRAS.329..513D,2004MNRAS.352..721E} 
based on the $\lambda_R$-parameter ($\lambda_{\mathrm{R}} \gtrsim $ 0.1: fast rotator,  
$\lambda_{\mathrm{R}} <$ 0.1: slow rotator), which gives an approximate measure
of the specific angular momentum of galaxies from their two-dimensional
line-of-sight velocity field \citep{2007MNRAS.379..401E}. This
definition was improved by $ATLAS^{\rm{3D}}$ due to the larger sample
size \citep{2011MNRAS.414..888E}. Fast
rotators dominate the low- and intermediate-mass field population
\citep{2011MNRAS.416.1680C} and form a quite homogeneous family of
flattened, oblate systems with regular velocity fields. Typical slow
rotators dominate in high density environments, are among the most
massive and round galaxies, and have peculiar properties such as
kinematic twists and kinematically decoupled  components
\citep{2011MNRAS.414.2923K,2011MNRAS.414..888E}. The results
from $ATLAS^{\rm{3D}}$ demonstrate the power of two-dimensional
integral field measurements which complements our understanding of
the properties and the origins of early-type galaxies. A comprehensive
summary of the current state, but without including integral field
kinmemtatics, can be found in \citet{2009ApJS..182..216K} and 
\citet{2012ApJS..198....2K}.      

It is the aim of the present study to take the next step, include the
observed two-dimensional kinematics, and investigate how the wealth of
kinematic features observed in nearby elliptical galaxies compares to
high resolution simulations  of massive galaxies evolving in a full
$\Lambda$CDM context. We also investigate how the results compare to
previous results which are mainly based on idealized major merger
simulations. For this purpose we present the first detailed
two-dimensional kinematic  analysis of a large sample of  44
individual zoom simulations of galaxies. To put our simulations in a
proper theoretical context we  first give a brief overview of the
information on the formation and evolution scenarios from previously
published studies (section \ref{theory}). In section \ref{simulations}
we describe the numerical simulations of the galaxies which are used
to construct the two-dimensional kinematic maps (section
\ref{maps}). The detailed properties and the origin of the kinematic
features are presented in Section \ref{linking} followed by a
discussion of projection effects and satellite properties (Section 
\ref{projection}). We then summarise and discuss our results in
Section \ref{summary}.  

\section{The theoretical context}
\label{theory}

Interestingly, one of the first theories, even before the
advent of modern hierarchical cosmologies, was motivated 
by what were then new measurements of the kinematics of stars  
in the outer halo of the Milky Way. The radial orbits of old,
non-rotating, halo stars were explained by their rapid formation  
in collapsing protogalactic cold gas streams, followed by the
formation of stars on circular orbits in a thin cold gas disc
\citep{1962ApJ...136..748E}. If this simple formation scenario were
similar for all old spheroidal systems they would consist of a spherical
bulge with stars on radial orbits and a rotating disc component
whose relative mass would depend on the efficiency of star formation
(i.e. gas consumption) during the collapse and on the ability of the gas
to cool. The amount of “turbulent viscosity”, caused by large
inhomogeneities in the infalling gas clouds, and the angular momentum
of the infalling gas would then determine the predicted galaxy properties
such as isophotal shapes, rotation, age and metallicity gradients, and even
the formation of disc-like substructures in elliptical galaxies 
\citep{1967ApJ...147..868P,1969MNRAS.145..405L,1974MNRAS.169..229L,1973ApJ...179..427S,1975MNRAS.173..671L}. In 
the pre-merger-scenario and pre-$\Lambda$CDM model for
the formation of slow and fast rotators the bulk of 
the stars in slow rotators would form in rapidly collapsing gaseous systems
with efficient star formation and efficient gas heating. Fast rotators
would form in more settled systems with inefficient star formation in the
absence of strong heating processes.             

An alternative (or supplementary) scenario was provided by 
\citet{1972ApJ...178..623T} and
\citet{1974IAUS...58..347T,1977egsp.conf..401T} who investigated
tidal interactions between disc galaxies and the possible formation of 
spheroidal systems by the morphological transformation of disc
galaxies in major mergers, rare but spectacular events that can be
observed to large distances. This scenario  became particularly
attractive as mergers play an important role during the 
formation and/or evolution of every dark matter halo and almost every 
massive galaxy in modern hierarchical cosmological models
\citep{1978MNRAS.183..341W,2006ApJ...648L..21K,2007MNRAS.375....2D}. The
first merger simulations
triggered a whole industry of simulations of mergers of discs with discs
or spheroids with spheroids. It is worth noting
that a significant fraction of the theoretical understanding on the
formation and evolution of early-type galaxies (kinematic
features in particular, which is the basis for this study also) is largely based on results from 
idealised merger simulations and it is therefore
necessary to briefly summarise the most important findings. 

Early collisionless simulations of the evolution of already existing
elliptical galaxies by mergers with other spheroids were mainly used to
asses the evolution of abundance gradients, shapes and kinematics
\citep{1978MNRAS.184..185W,1979MNRAS.189..831W}. Later 
investigations of spheroid mergers tested in more detail the evolution of
scaling relations, also including the effect of different merger mass-ratios
\citep{1997ApJ...481...83M,2003MNRAS.342..501N,2005MNRAS.362..184B,2006ApJ...636L..81N,2006MNRAS.369.1081B,2008MNRAS.383...93B,2009ApJ...703.1531N,2009ApJ...706L..86N,2009A&A...501L...9D,2009ApJ...703.1531N,2012MNRAS.422.1714N}.
These studies received particular attention since it became clear
that major collisionless 'dry' stellar mergers actually happen in the
Universe, i.e. they can be directly observed in significant numbers up 
to high redshifts
\citep{2005AJ....130.2647V,2005ApJ...627L..25T,2006ApJ...640..241B,2006ApJ...652..270B,2008ApJ...672..177L,2008MNRAS.388.1537M,2009ApJ...697.1971J,2012ApJ...747...85X,2012ApJ...747...34B,2012ApJ...746..162N,2012ApJ...744...85M}   
in combination with the fact that the late mass assembly of massive
galaxies by collisionless mergers is theoretically expected in
hierarchical cosmological models 
\citep{1996MNRAS.283L.117K,1996MNRAS.281..487K,2003ApJ...597L.117K,2006MNRAS.366..499D,2006ApJ...648L..21K,2007MNRAS.375....2D,2008MNRAS.384....2G,2009ApJS..182..216K,2010ApJ...715..202H}. The
terms collisionless or 'dry' are of course, an over-simplification as
they imply the absence of gas in these system. In reality, gas
is present in almost every galaxy but sometimes it can be considered
dynamcially unimportant \citep{2007AJ....134.1118D,2009MNRAS.400.1264S,2010MNRAS.401L..29S}. 

However, the difficulty of forming the observed family of slowly
rotating early-type galaxies in major mergers of spheroids was pointed out
early-on \citep{1979ApJ...229L...9W}, a problem that has been confirmed with
modern high-resolution simulations. In general, collisionless major
mergers produce remnants that are either fast rotating - even if the
progenitors did not rotate - or are intrinsically too flattened to be
consistent with observed slow rotators
\citep{2009A&A...501L...9D,2010MNRAS.406.2405B,2011MNRAS.416.1654B}.             

\begin{figure*}
\begin{center}
  \epsfig{file=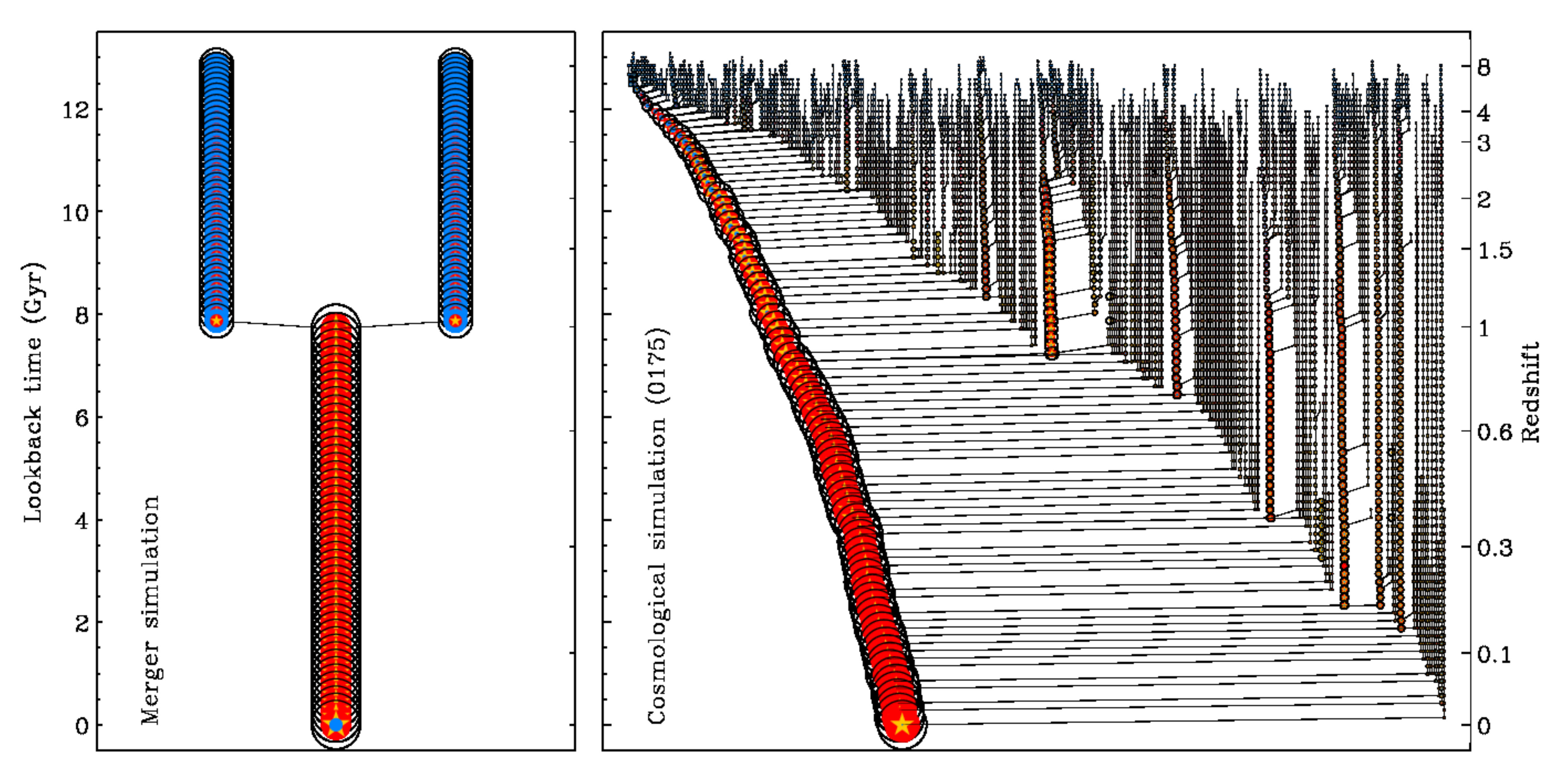, width=\textwidth}
  \caption{{\it Left panel}: Schematic 'merger-tree' representation of
    a binary disc merger simulation. Two gas-rich (blue) stellar
    (yellow) discs with little hot gas (red) merge at $z \approx 1$ and
    form an elliptical galaxy. {\it Right panel}: Merger tree from a
    cosmological zoom simulation of the formation of a halo (0175) and 
    its galaxy within the concordance cosmology. Black circles
    indicate the dark matter halo mass at every redshift with the
    symbol size 
    proportional to the square root of the normalized halo mass at $z
    = 0$. The yellow stars indicate stellar mass, the blue and red
    filled circles show the cold and hot gas mass within the virial
    radius \citep{2012MNRAS.419.3200H}. It is evident that
    continuous infall of matter in small and large units is an
    important characteristic of the assembly of massive galaxies (see  
    e.g. \citealp{2007MNRAS.375....2D}). The galaxy shown (0175) has no 
    major merger since $z \approx 3$. Others galaxies of similar mass
    can have up to three major mergers. The mass growth, however, is always
    accompanied by minor mergers (see Section \ref{linking}).}    
\label{TreeVis}
\end{center}
\end{figure*}

A separate process that was investigated using idealized merger
simulations was the formation of elliptical galaxies by the 
morphological transformation of disc galaxies in major spiral galaxy
mergers. The advantage of this idealized approach was the ability to
study fundamental dynamical and gas-dynamical processes at high
spatial resolution with well controlled initial conditions. 
The first self-consistent disc merger simulations clearly
demonstrated that major mergers can indeed transform kinematically
cold discs into a kinematically hot spheroid with elliptical-like
properties
\citep{1981MNRAS.197..179G,1982ApJ...259..103F,1983MNRAS.205.1009N,1988ApJ...331..699B,1992ARA&A..30..705B,1992ApJ...400..460H,1992ApJ...393..484B,1993ApJ...409..548H,1994ApJ...427..165H}. Here
the mass-ratio has a significant impact on the remnant properties 
with remnants of more unequal-mass mergers showing faster rotation and
being more flattened with more discy isophotes
\citep{1998giis.conf..275B,1998ApJ...502L.133B,1999ApJ...523L.133N,2000MNRAS.316..315B,1999ApJ...523L.133N,2003ApJ...597..893N,2004A&A...418L..27B,2005A&A...437...69B,2005MNRAS.360.1185J,2005MNRAS.357..753G,2005A&A...437...69B,2006MNRAS.369..625N,2007A&A...476.1179B,2009MNRAS.397.1202J}.    
Mergers also have a dramatic impact on the extended gas
components in the progenitor discs.  The gas is torqued, loses its
angular momentum and is driven to the central regions. Therefore
mergers were early-on considered to trigger starbursts as observed in 
local interacting galaxies
\citep{1994ApJ...431L...9M,1996ApJ...464..641M,2004MNRAS.350..798B,2007A&A...468...61D,2008A&A...492...31D,2010ApJ...715L..88K,2010ApJ...720L.149T}  
and eventually feed central super-massive black holes 
\citep{1989Natur.340..687H,2005MNRAS.361..776S,2005Natur.433..604D,2009ApJ...690..802J,2009MNRAS.396L..66Y}. In 
addition, it was shown originally by \citet{1996ApJ...471..115B} that a
dissipational component has a significant impact on the properties of
the stellar remnant. In general, gas makes the centers of the remnants
rounder, less boxy, more centrally concentrated with well studied and
plausible effects on the star formation efficiencies and scaling
relations
\citep{1994ApJ...437L..47M,1997ApJ...478L..17B,2000MNRAS.312..859S,2006ApJ...641...21R,2006MNRAS.370.1445D,2006ApJ...650..791C,2008ApJ...679..156H,2009ApJS..181..135H,2009ApJS..181..486H,2009ApJ...691.1424H,2009ApJ...690..802J,2010MNRAS.406L..55D,2011MNRAS.415.3750M}.   
Inflowing gas also changes the shape of the central potential which determines the
availability of orbital families for the stars 
\citep{1996ApJ...471..115B,2005MNRAS.360.1185J,2006MNRAS.372L..78G,2006MNRAS.372..839N,2009ApJ...705..920H,2010ApJ...723..818H}. It
was shown by \citet{2006MNRAS.372..839N} that the presence of a
dissipational component changes the asymmetry of the observable
line-of-sight velocity distributions towards steep leading wings
in agreement with observed rotating early-type galaxies.

\citet{2000MNRAS.316..315B} presented the first two-dimensional analysis of
line-of-sight velocity distributions of simulated disc mergers
remnants for a direct comparison with observational results from 
integral field spectroscopy. For equal-mass merger remnants they
identified a variety of observed kinematic features such as 
counter-rotating cores and misaligned rotation. Unequal-mass merger
remnants  on the other hand showed relatively regular rotation. This
was confirmed by the first quantitative studies by
\citet{2007MNRAS.376..997J,2009MNRAS.397.1202J} who compared the kinematic 
features of a sample of simulated remnants to results from the SAURON
survey. Interestingly, depending only on the relative initial
orientation of the progenitor discs, almost every observed kinematic peculiarity was
found in one of the equal-mass merger remnants including major axis
rotation, kinematic twists, dumbbell features (a feature in the
stellar dispersion) and counter-rotating
cores which have long been considered to originate from galaxy mergers  
\citep{1990ApJ...361..381B,1991Natur.354..210H,2008A&A...477..437D,2010MNRAS.406.2405B}.   
>From the studies of \citet{2009MNRAS.397.1202J}, 
\citet{2010MNRAS.406.2405B}, and \citet{2011MNRAS.416.1654B} it became clear that
only major disc mergers - depending on the initial disc orientation -
can form slow rotators which mostly have counter rotating
cores. Still, the problem remains that most of the slow rotators
formed in this way are too flat to be consistent with observed slowly
rotating early-type galaxies. The kinematics of observed fast rotating
early-type galaxies, on the other hand, seems to be in very good agreement
with remnants of 'minor' disc mergers with varying mass-ratios
\citep{2009MNRAS.397.1202J,2011MNRAS.416.1654B}.

Despite the apparently overwhelming successes of binary merger
simulations in explaining photometric and kinematic properties of
early-type galaxies this approach comprises considerable
limitations. Simulations of mergers of bulges only address aspects of
early-type galaxy evolution and not their formation. Mergers of disc galaxies do
form new spheroids but in general the stellar populations of present
day disc galaxies and their progenitors have too low masses and are
too young and too metal poor to account for the bulk of the present
day massive early-type galaxy population
\citep{2009ApJ...690.1452N}. Even more striking is the fact that
assembly histories of massive galaxies in currently favored
hierarchical cosmological models are significantly more complex than a
single binary merger. They grow   
- in particular at high redshift - by smooth accretion of gas, major
mergers but also numerous minor mergers covering a large range of
mass-ratios which can dominate the mass of assembled stars
\citep{2007MNRAS.375....2D,2008ApJ...688..789G,2010ApJ...709..218F,2010ApJ...725.2312O,2011ApJ...736...88F,2012MNRAS.419.3200H}. This
conceptual difference is illustrated in Fig. \ref{TreeVis} where
we represent in the left panel the idealized  assembly 
history of a binary disc merger by 
an artificially constructed merger tree. At some redshift which is
high enough so that the stellar population has enough time to age and
become red a merger turns two 'blue' disc galaxies into a 'red and dead'
elliptical. In the right panel we show the complete merger tree of a
cosmological hydrodynamical zoom simulation
\citep{2010ApJ...725.2312O} of a massive galaxy (see
\citet{2012MNRAS.419.3200H} for details on the construction of the
tree). The differences are obvious. 

The picture emerging from recent high-resolution cosmological
zoom-in simulations on the formation of massive galaxies comprises
characteristic features of both the 'monolithic' dissipative collapse
models of the early 70's and the merger scenario of the 80's and
90's. Massive early-type galaxies appear to grow in two main
phases. The early assembly ($2 < z < 6$) is dominated by significant
gas inflows \citep{2005MNRAS.363....2K,2009Natur.457..451D} and the in-situ
formation of stars whereas the late evolution is dominated by the
assembly of stars which have formed in other galaxies and have then been
accreted onto the system at lower redshifts ($3 < z < 0$) 
\citep{2003ApJ...590..619M,2007ApJ...658..710N,2008ApJ...684.1062F,2009ApJ...699L.178N,2010ApJ...709..218F,2010ApJ...725.2312O,2011ApJ...736...88F,2012arXiv1202.3441J,2012arXiv1206.0295L}. So
far it has been demonstrated that properties of the spheroidal galaxies
from cosmological simulations are in reasonable global agreement with
observations of the cosmological evolution of scaling relations and
present day properties of early-type galaxies
\citep{2007ApJ...658..710N,2009ApJ...699L.178N,2011ApJ...736...88F,2012arXiv1202.3441J,2012ApJ...744...63O}. 
However, a more detailed kinematic analysis was presented only for a few fast rotating low mass cases
revealing e.g. the presence of discy isophotes
and line-of-sight velocity distributions with steep leading wings 
\citep{2003ApJ...590..619M,2007ApJ...658..710N}. In this paper we
will close this gap and present a detailed two-dimensional analysis
for a much larger galaxy sample also covering galaxies with significantly higher stellar
masses. 

\section{High-resolution cosmological simulations of individual galaxies}
\label{simulations}
The analysis presented in this paper is based on 44 'zoom-in'
cosmological hydrodynamic simulations of individual massive galaxy
halos selected from a sample presented in
\citet{2010ApJ...725.2312O,2012ApJ...744...63O}. The halos considered
for re-simulation were chosen from a $512^3$ particle dark matter only
simulation of a $100^3 \ \rm Mpc^3$ volume using a WMAP3 cosmology
\citep{2007ApJS..170..377S}: $h = 0.72, \Omega_{b} = 0.044,
\Omega_{\mathrm{dm}} = 0.216, \Omega_{\Lambda} = 0.74, \sigma_8 =
0.77$, and an initial slope of the power spectrum of $n_s =
0.95$. More details of the parent simulation are presented in
\citet{2010ApJ...710..903M} and
\citet{2010ApJ...725.2312O,2012ApJ...744...63O} but we briefly review
the simulation setup relevant for the galaxy re-simulations presented
here.  All particles that are inside a sphere with radius $2 \times
R_{200}$ centered on a halo of interest in any of our 95 snapshots
were identified. We then selected a coherent convex volume in the
initial conditions that contains all those particles and replaced them
by high-resolution gas and dark matter particles including the
relevant small scale fluctuations. The original dark matter
distribution at larger radii was down-sampled to provide the proper
tidal field at a low computational cost. These initial conditions were
then evolved from redshift $z = 43$ to $z=0$ using the parallel
TreeSPH code GADGET \citep{2005MNRAS.364.1105S} including star
formation, supernova feedback \citep{2003MNRAS.339..289S} and cooling
for a primordial composition of hydrogen and helium. The simulations
also include a redshift dependent UV background radiation field with a
modified \citet{1996ApJ...461...20H} spectrum.

Here we consider a sample of massive halos with masses in the range of
$2.2 \times 10^{11} M_{\odot}$ $ \lesssim
M_{\mathrm{vir}} \lesssim 3.7 \times 10^{13} M_{\odot}$. The
halos host massive galaxies with present day stellar masses between
$2.6 \times 10^{10} M_{\odot} \lesssim M_* \lesssim 5.7 \times
10^{11} M_{\odot}$ and projected half-mass radii of $1.2 <
R_{1/2} < 6.6 \ \rm kpc$. The host halos were randomly chosen to evenly
cover the above halo mass range. It is therefore important to note
that the simulated  galaxies do not represent a statistically
significant sample of the whole galaxy populations in this mass 
range. It is beyond the scope of this paper to address the galaxy
properties as a population. We rather investigate the possible
variations in individual formation histories and global trends with mass. 

All galaxies are well resolved with $\approx 1.4
\times 10^4 - 2 \times 10^6$ particles within the virial radius. The
masses of individual gas and star particles are $m_{*,gas}=4.2 \times
10^{6}M_{\odot}$ (one star particle per gas particle is
spawned), and the dark matter particles have a mass of
$m_{\mathrm{dm}} = 2.5 \times 10^{7}M_{\odot}$. The comoving
gravitational softening lengths are $\epsilon_{\mathrm{gas,star}} =
400 \rm pc \, h^{-1}$ for gas and star particles and
$\epsilon_{\mathrm{halo}} = 890 \rm pc \, h^{-1}$ for dark matter
particles. This guarantees that  the simulated
half-mass radii of all galaxies presented here are well resolved (see
e.g. \citealp{2010ApJ...717..121C})

In Tab. \ref{tab1} we list some of the basic properties of the
central galaxies that we find in these simulations. We use SUBFIND
\citep{2001MNRAS.328..726S} to identify the central galaxies and
determine stellar masses and projected half-mass radii inside of $0.1
\times R_{200}$. This cut-off may yield slightly lower radii than in
observed galaxies of the same mass but it provides us with a
well-defined fiducial value for galaxies that are spread over more than an
order of magnitude in mass. 

In order to investigate the assembly histories of the central galaxies we
identify every satellite inside $0.15 \times R_{200}$ using a
friends-of-friends (FOF) finder with a minimum number of 20 stellar particles ($\approx 1.2
\times 10^{8} M_{\odot}$, i.e. for all the galaxies in our sample we
can identify mergers down to a mass-ratio of 1:5 for
redshifts below 2) at which time we set the mass-ratio of the
merger. As the point in time when the merger actually takes place we 
use the first snapshot in which the most bound stellar particle of the
satellite can be found in the same FOF group as the central galaxy.
This allows us to compute an average number weighted merger ratio for
all merger events between redshift of 2 and the present day
(NWMR). Since all galaxies encounter many minor mergers which
do not necessarily account for the majority of the final accreted mass
we also define the mass-weighted merger ratio (MWMR) where every merger
ratio is weighted with the stellar mass of the accreted system. 

The additional present day parameters presented in Table \ref{tab1} are the galaxy mass
($M_{\star}$) which we define as the stellar mass enclosed in $0.1
\times R_{200}$, the edge-on projected half-mass radius $R_{1/2}$, and
the projected ellipticity at $R_{1/2}$. Furthermore we calculate the
three dimensional moment-of-inertia tensor for all particles inside
$R_{1/2}$ and determine the ratios of its principle axes ($b/a, c/a$).
As a measure for the importance of dissipation in the formation of the 
galaxies we present the ratio of stars that formed in-situ to the total
stellar mass at the present day ($M_{ins}/M_{\star}$). Finally we
quantify the rotation of the halo with the dimensionless spin
parameter $\lambda_{H}$ as defined in \cite{2001ApJ...555..240B}.


\clearpage 

\begin{deluxetable}{rrrcccrrrrrrrrrrrrrr}
\tablewidth{0pt}
\tablecaption{Table Caption\label{tab1}}
\tablehead{
 \colhead{ID} &
 \colhead{$M_{\mathrm{vir}}$} & 
 \colhead{$M_{*}$} &
 \colhead{$R_{1/2}$} &
 \colhead{$\lambda_{\mathrm{R}}$} &
 \colhead{$\epsilon$} &
 \colhead{q} &
 \colhead{s} &
 \colhead{$M_{ins}/M_{*}$} &
 \colhead{$\lambda_H$} &
 \colhead{MWMR} &
 \colhead{NWMR} &
 \colhead{Class} & \\
 \colhead{(1)} &
 \colhead{(2)} &
 \colhead{(3)} &
 \colhead{(4)} &
 \colhead{(5)} &
 \colhead{(6)} &
 \colhead{(7)} &
 \colhead{(8)} &
 \colhead{(9)} &
 \colhead{(10)} &
 \colhead{(11)} &
 \colhead{(12)} &
 \colhead{(13)} 
}
\startdata
M0040 &   3716  &         42.42 &         6.55 &        0.13 &        0.36 &        0.87 &        0.80 &         0.11 &      0.053 &        0.30 &      0.038 & D \\
M0053 &   2327  &         56.89 &         6.36 &      0.093 &        0.44 &        0.87 &        0.74 &         0.19 &      0.013 &        0.11 &      0.020 & E \\
M0069 &   2466  &         41.72 &         4.56 &        0.15 &        0.41 &        0.81 &        0.70 &         0.15 &      0.082 &        0.18 &      0.032 & E \\
M0089 &   1478  &         38.97 &         4.65 &      0.074 &        0.44 &        0.80 &        0.70 &       0.09 &      0.023 &        0.19 &      0.042 & E \\
M0094 &   1394  &         42.11 &         4.13 &      0.098 &        0.41 &        0.84 &        0.71 &         0.16 &      0.034 &        0.29 &      0.029 & E \\
M0125 &   1273  &         38.23 &         5.40 &      0.078 &        0.27 &        0.85 &        0.83 &         0.12 &      0.035 &        0.11 &      0.021 & F \\
M0162 &   875.1  &         28.13 &         4.32 &      0.074 &        0.53 &        0.71 &        0.62 &       0.081 &      0.039 &        0.59 &      0.072 & E \\
M0163 &   956.3  &         27.90 &         4.86 &        0.31 &        0.44 &        0.80 &        0.72 &       0.098 &      0.028 &        0.22 &      0.046 & D \\
M0175 &   970.2  &         32.57 &         4.28 &      0.058 &        0.31 &        0.89 &        0.79 &         0.14 &      0.043 &      0.076 &      0.020 & F \\
M0190 &   709.3  &         27.74 &         4.09 &      0.083 &        0.56 &        0.70 &        0.59 &       0.093 &      0.045 &        0.41 &      0.072 & E \\
M0204 &   746.5  &         23.57 &         3.61 &      0.100 &        0.23 &        0.94 &        0.84 &         0.12 &      0.046 &        0.25 &      0.056 & F \\
M0209 &   826.4  &         18.54 &         2.49 &        0.14 &        0.40 &        0.92 &        0.72 &         0.17 &      0.079 &        0.17 &      0.038 & E \\
M0215 &   701.0  &         24.79 &         3.31 &        0.14 &        0.28 &        0.91 &        0.79 &         0.16 &      0.028 &        0.11 &      0.027 & E \\
M0224 &   663.8  &         21.40 &         3.01 &        0.16 &        0.35 &        0.95 &        0.75 &         0.14 &      0.040 &        0.26 &      0.053 & D \\
M0227 &   707.0  &         26.61 &         4.07 &        0.24 &        0.33 &        0.88 &        0.76 &         0.10 &      0.057 &        0.31 &      0.047 & D \\
M0259 &   606.5  &         17.93 &         2.81 &        0.40 &        0.40 &        0.99 &        0.70 &         0.15 &      0.045 &        0.34 &      0.049 & A \\
M0290 &   581.7  &         19.78 &         2.05 &        0.48 &        0.30 &        0.97 &        0.75 &         0.19 &      0.036 &        0.22 &      0.062 & B \\
M0300 &   507.3  &         17.01 &         2.72 &        0.19 &        0.50 &        0.90 &        0.64 &         0.12 &      0.069 &        0.21 &      0.056 & E \\
M0329 &   486.5  &         19.70 &         2.96 &      0.071 &        0.34 &        0.90 &        0.79 &         0.16 &      0.033 &        0.10 &      0.026 & F \\
M0380 &   455.8  &         15.72 &         2.61 &        0.46 &        0.42 &        0.84 &        0.68 &         0.17 &      0.038 &      0.021 &      0.014 & A \\
M0408 &   350.8  &         16.36 &         2.13 &        0.37 &        0.38 &        0.98 &        0.73 &         0.20 &      0.047 &        0.22 &      0.063 & B \\
M0443 &   371.7  &         21.30 &         1.95 &      0.088 &        0.32 &        0.91 &        0.77 &         0.24 &      0.027 &        0.20 &      0.035 & C \\
M0501 &   319.4  &         13.70 &         2.46 &      0.075 &        0.38 &        0.87 &        0.72 &         0.16 &      0.051 &        0.11 &      0.030 & E \\
M0549 &   300.1  &         10.13 &         2.53 &        0.46 &        0.35 &        0.93 &        0.71 &         0.17 &      0.056 &      0.074 &      0.032 & A \\
M0616 &   262.9  &         12.13 &         2.95 &      0.077 &        0.40 &        0.84 &        0.75 &         0.16 &      0.040 &        0.11 &      0.044 & E \\
M0664 &   249.1  &          9.58 &         2.03 &        0.14 &        0.30 &        0.87 &        0.78 &         0.17 &      0.012 &        0.13 &      0.043 & C \\
M0721 &   204.7  &         12.69 &         1.64 &        0.40 &        0.50 &        0.97 &        0.71 &         0.38 &      0.043 &        0.24 &      0.075 & A \\
M0763 &   208.0  &         11.77 &         2.34 &        0.32 &        0.34 &        0.91 &        0.74 &       0.099 &      0.066 &        0.39 &        0.13 & D \\
M0858 &   193.1  &         13.48 &         1.94 &        0.52 &        0.51 &        0.90 &        0.65 &         0.32 &      0.024 &        0.29 &        0.15 & A \\
M0908 &   173.1  &         12.90 &         1.96 &        0.44 &        0.44 &        0.98 &        0.70 &         0.38 &      0.013 &        0.41 &        0.20 & A \\
M0948 &   167.9  &          7.77 &         2.83 &      0.088 &        0.19 &        0.94 &        0.88 &         0.12 &      0.014 &      0.099 &      0.045 & F \\
M0959 &   166.6  &          7.79 &         2.13 &      0.091 &        0.25 &        0.83 &        0.81 &         0.18 &      0.040 &        0.15 &      0.043 & C \\
M0977 &   131.0  &          5.63 &         1.93 &        0.35 &        0.53 &        0.91 &        0.66 &         0.37 &      0.085 &        0.24 &      0.095 & B \\
M1017 &   147.1  &          8.49 &         1.68 &      0.084 &        0.43 &        0.86 &        0.71 &         0.29 &      0.017 &        0.10 &      0.046 & C \\
M1071 &   146.9  &         10.09 &         1.69 &        0.14 &        0.20 &        0.95 &        0.85 &         0.21 &     0.008 &        0.19 &      0.081 & C \\
M1167 &   129.1  &          9.38 &         1.59 &      0.062 &        0.40 &        0.84 &        0.69 &         0.26 &      0.027 &        0.36 &        0.14 & C \\
M1192 &   108.3  &          5.53 &         1.87 &        0.50 &        0.52 &        0.75 &        0.58 &         0.20 &      0.038 &        0.17 &      0.086 & A \\
M1196 &   132.4  &         10.27 &         2.04 &        0.45 &        0.49 &        0.97 &        0.69 &         0.33 &      0.048 &        0.20 &      0.056 & B \\
M1306 &   108.8  &          8.53 &         1.35 &        0.59 &        0.39 &        0.96 &        0.65 &         0.28 &      0.024 &        0.16 &      0.075 & A \\
M1646 &   99.01  &          7.20 &         1.86 &        0.28 &        0.52 &        0.99 &        0.68 &         0.35 &      0.015 &      0.092 &      0.034 & A \\
M2665 &   53.90  &          4.18 &         1.57 &      0.098 &        0.26 &        0.89 &        0.83 &         0.29 &      0.028 &        0.21 &        0.11 & C \\
M3852 &   42.33  &          3.49 &         1.21 &        0.57 &        0.42 &        0.96 &        0.63 &         0.29 &      0.092 &        0.35 &        0.13 & B \\
M5014 &   32.46  &          3.00 &         1.22 &        0.51 &        0.46 &        0.98 &        0.67 &         0.38 &      0.069 &        0.36 &        0.41 & B \\
M6782 &   21.97  &          2.55 &         1.21 &        0.55 &        0.54 &        0.84 &        0.52 &         0.27 &      0.077 &        0.85 &        0.48 & B \\
\enddata
\tablecomments{
(1): ID of the galaxy.
(2): virial Mass in $10^{10}M_{\odot}$
(3): Stellar mass inside $r_{10}$ in $10^{10}M_{\odot}$
(4): Projected half-mass radius kpc
(5): $\lambda_{\mathrm{R}}$-parameter
(6): Projected ellipticity at $R_{1/2}$
(7): Intermediate to major axis ratio
(8): Minor to major axis ratio
(9): Ratio of in-situ formed to total stellar mass
(10): Halo spin parameter
(11): Mass-weighted merger ratio
(12): Number-weighted merger ratio
(13): Assembly class (see Section \ref{linking})
}
\end{deluxetable}
\clearpage

\section{Construction of two-dimensional kinematic maps and data
  analysis} 
\label{maps}

Until now two-dimensional velocity fields from numerical galaxy
simulations have only been constructed and analysed for remnants of
binary merger simulations which provided the necessary resolution for a
reliable analysis   
\citep{2000MNRAS.316..315B,2007MNRAS.376..997J,2009MNRAS.397.1202J,2009ApJ...705..920H,2010ApJ...723..818H,2010MNRAS.406.2405B,2011MNRAS.416.1654B}. The
high spatial and mass resolution of modern cosmological simulations of
individual galaxies makes it now possible to extend the
two-dimensional analysis to simulated galaxies that form and evolve in
a full cosmological context. 

The two-dimensional kinematic maps presented here are constructed in a
similar way as  described in
\citet{2007MNRAS.376..997J,2009MNRAS.397.1202J}, with a  
few notable differences in order to follow as closely as possible the data
analysis for the galaxies in the ATLAS$^{\rm{3D}}$ sample. In a 
first step we identify the main galaxy of each simulated halo and
shift all positions and velocities to its baryonic center 
using a shrinking sphere technique. The stellar component of the
galaxy is rotated according to the principal axes of the
moment-of-inertia tensor of the 50 per cent most tightly bound stellar 
particles. As a reference measure and for comparison with the directly
observable stellar half-light radius, or effective radius
$r_{\mathrm{e}}$, we compute the edge-on projected
(along the minor-axis) circular stellar half-mass radius, $R_{1/2}$, within 10
percent of the virial radius of the galaxy
\citep{2012ApJ...744...63O}. For every projected stellar particle
residing in a box of two half-mass radii side length, centered on the
galaxy, we create a set of 60 pseudo-particles with
identical line-of-sight velocities and 1/60th the original particle
mass. The pseudo-particles are distributed in  
the plane of the sky following a two-dimensional Gaussian with a
standard deviation of 0.3 kpc. In this
way we account for seeing effects on the projected mass and velocity
distributions and the limited spatial resolution of the simulations.      

All pseudo-particles are then binned on a spatial grid, centered on
the projected particle position with four half-mass radii side length
and a pixel size of $200 \rm pc$. The grid has variable dimensions
depending on the projected size of the galaxy. Our chosen pixel size
approximately corresponds to the spatial coverage of one lens-let of
the SAURON instrument \citep{2001MNRAS.326...23B}
at a distance of $20 \rm Mpc$. In contrast to simulations, real
galaxies are observed with an instrument of fixed angular coverage and
the spatial coverage varies between 0.5 and 3 effective radii
depending on the physical size and distance of the galaxies
\citep{2011MNRAS.413..813C}. 
  
Using the regularly binned spatial data we group, whenever necessary, 
adjacent bins into larger bins with a comparable pre-defined
signal-to-noise ratio using a Voronoi tessellation method as described
in \citet{2003MNRAS.342..345C}. This results in an irregular grid
structure but guarantees that all bins contain approximately the same
number of particles. 
>From the velocity data we construct line-of-sight velocity profiles for each
Voronoi bin along the two-dimensional grid. 

\begin{figure*}
\begin{center}
  \epsfig{file=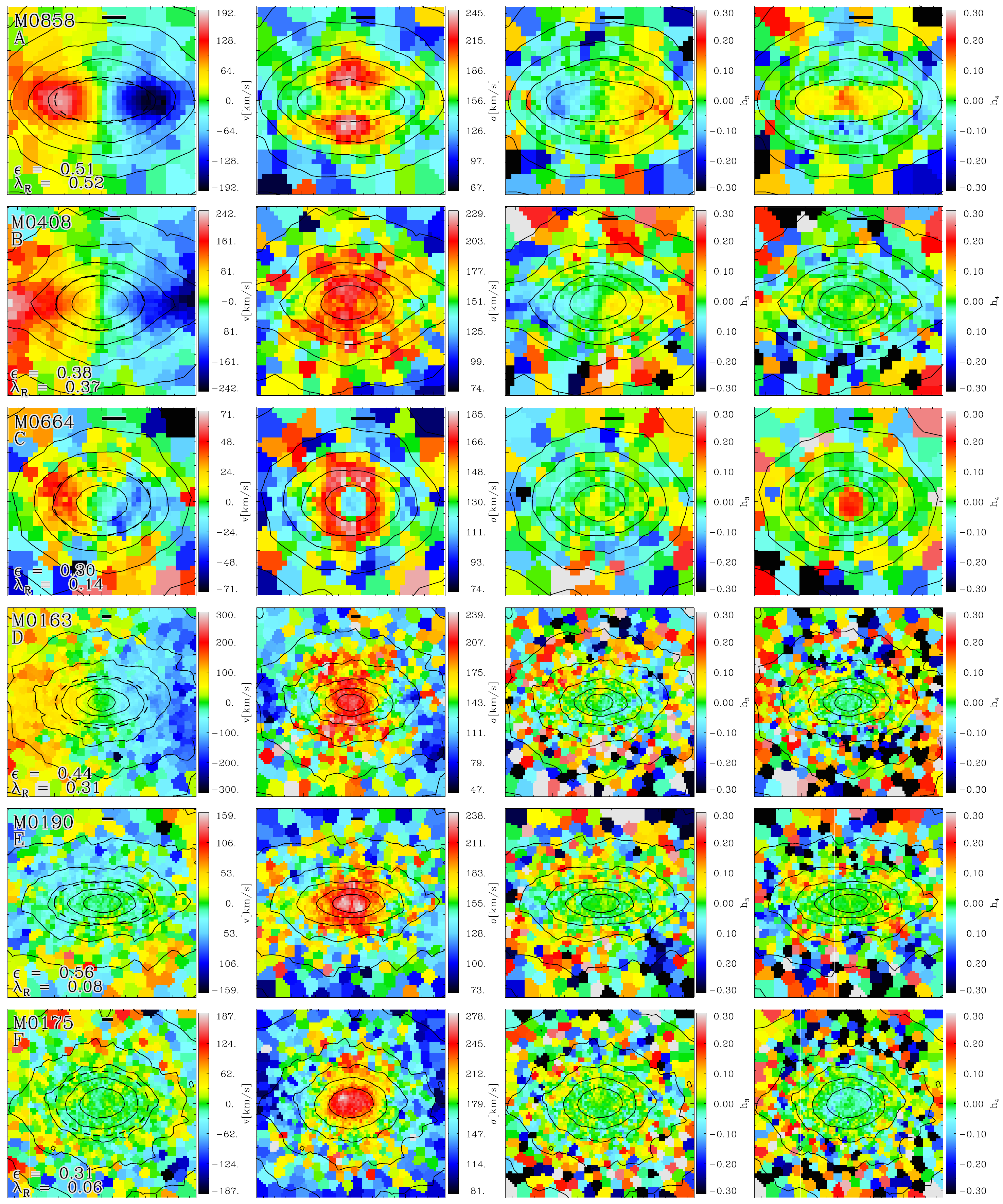,width=0.9\textwidth}
  \caption{Two-dimensional maps of the stellar kinematics for edge-on
    projections of six simulated galaxies M0858, M0408, M0664, M0163,
    and M0190, and M0175 (from top to
    bottom). These galaxies 
    are representatives of galaxy classes (discussed in the text) A: Fast
    rotators with central dissipation, B: Dissipative fast rotators,
    C: Dissipational slow rotators, D: Dissipationless fast rotators,
    E: Elongated dissipationless slow-rotators, and F:
    Dissipationless round slow rotators. The box length is three
    projected stellar half-mass radii ($R_{1/2}$). The contour lines
    show the projected stellar mass surface density. The galaxy ID,
    the $\lambda_{\mathrm{R}}$ parameter, the ellipticity at
    $R_{1/2}$,  and the physical scale of $1 \ \rm kpc$ (black bar) are given 
    in the leftmost panels. The $\lambda_{\mathrm{R}}$ parameter is
    calculated within the effective elliptical radius (dotted
    line). From the left to the right we show
    line-of-sight velocity ($v$), line-of-sight velocity dispersion
    ($\sigma$), a measure for the asymmetric ($h_3$), and 
    symmetric ($h_4$) deviations from a Gaussian LOSVD shape. The maps
    of all simulated galaxies can be found in Appendix \ref{samplemaps}. }
\label{gallery-extra}
\end{center}
\end{figure*}

To get a quantitative measurement for the deviations of the LOSVD from 
the Gaussian shape the velocity profile $P(v)$  can be parameterized in
accordance with \citet{1993MNRAS.265..213G} and
\citet{1993ApJ...407..525V} by a Gaussian plus third- and fourth-order 
Gauss-Hermite functions (see \citealp{1994MNRAS.269..785B}).  
The third and fourth order amplitudes $h_3$ and 
$h_4$ are related to the skewness and the
kurtosis of the velocity profile but the skewness/kurtosis and
$h_3$/$h_4$ are not identical. The skewness and 
kurtosis are the normalized third- and fourth-order moments of the
LOSVD and are more susceptible to the wings of the line profile 
which are ill-constrained by the observations
(see \citealp{1993ApJ...407..525V}), here  $\gamma$ is a normalization
constant. If the LOSVDs deviate from a Gaussian, the fit parameters
$v_{\mathrm{fit}}$ and $\sigma_{\mathrm{fit}}$ correspond only to first order
to the real first ($v_{\mathrm{los}}$) and second
($\sigma_{\mathrm{los}}$) moment of the velocity distribution
(differences can be up to up to $15\%$, see \citealp{1994MNRAS.269..785B};
\citealp{1994MNRAS.271..949M}). For 
$h_3= 0$ and $h_4 = 0$ the velocity profile is a
Gaussian. The values of $h_3$ and $v_{\mathrm{fit}}$ have opposite
signs for asymmetric profiles with the pro-grade (leading) wing
being steeper than the retrograde (trailing) wing, which can be
indicative of an embedded disc structure 
\citep{1993ApJ...407..525V,1994MNRAS.269..785B,1997AJ....113..950F,2008MNRAS.390...93K}. 
When $v_{\mathrm{fit}}$ and $h_3$ have the same sign, the leading wing is broad
and the trailing wing is narrow. LOSVDs with $h_4 > 0$ have a
peaked shape, where the distribution's peak is narrow
with broad wings. Flat-top LOSVDs have $h_4 < 0$ where the peak is
broad and the wings are narrow. The kinematic parameters of the LOSVD
in each Voronoi bin ($v_{\mathrm{fit}}$, $\sigma_{\mathrm{fit}}$,
$h_3$, $h_4$) are determined from the discrete in a maximum likelyhood
as in \citet{2006A&A...445..513V} which is particularly suitable for our
purposes. The signal-to-noise value in the simulations is driven by
the particle number and the numerical noise is significantly reduced
by the pseudo-particle procedure. Whenever indicated we repeat the above analysis for
different projections of the galaxies on the plane of the sky.   

From the two-dimensional velocity maps we compute the
$\lambda_{\mathrm{R}}$-parameter as introduced by  \citet{2004MNRAS.352..721E}
according to   
\begin{equation}
\lambda_R=\frac{\Sigma_{i=1}^{N_p} F_i R_i
  |V_i|}{\Sigma_{i=1}^{N_p}F_i R_i \sqrt{V_i^2 +\sigma_i^2}}, 
\end{equation}
where $F_i$ is the flux (here the projected mass in every bin), $R_i$ 
the projected radius, $v_i$ the  line-of-sight velocity and $\sigma_i$
the line-of-sight velocity dispersion of each grid cell. We calculate
$\lambda_{\mathrm{R}}$ as a function of radius for every 
galaxy from its two-dimensional map \citep{2009MNRAS.397.1202J}. When
computing a characteristic value of $\lambda_{\mathrm{R}}$ we have to consider
that $ATLAS^{\rm{3D}}$ has a finite field-of-view which typically extends to
0.3 - 3 effective radii \citep{2011MNRAS.414..888E}. To take this into account we determine the
half-mass radius for each projection and sum only over the grid cells
inside one circular half-mass radius. As $\lambda_{\mathrm{R}}$ is a
cumulative parameter of absolute values the numerical noise results in
a lower limit for our our measurements of $\lambda_{\mathrm{R}} \sim
0.05$ (see \citealp{2012arXiv1209.3741W} for a detailed
discussion). This procedure ensure a fair comparison to the
$ATLAS^{\rm{3D}}$ data. The global values of $\lambda_{\mathrm{R}}$
for our galaxies are given in Table \ref{tab1} as well as in the left
panels of Figs. \ref{gallery-extra} and \ref{maps0} - \ref{maps7}.

\begin{figure*}
\begin{center}
  \epsfig{file=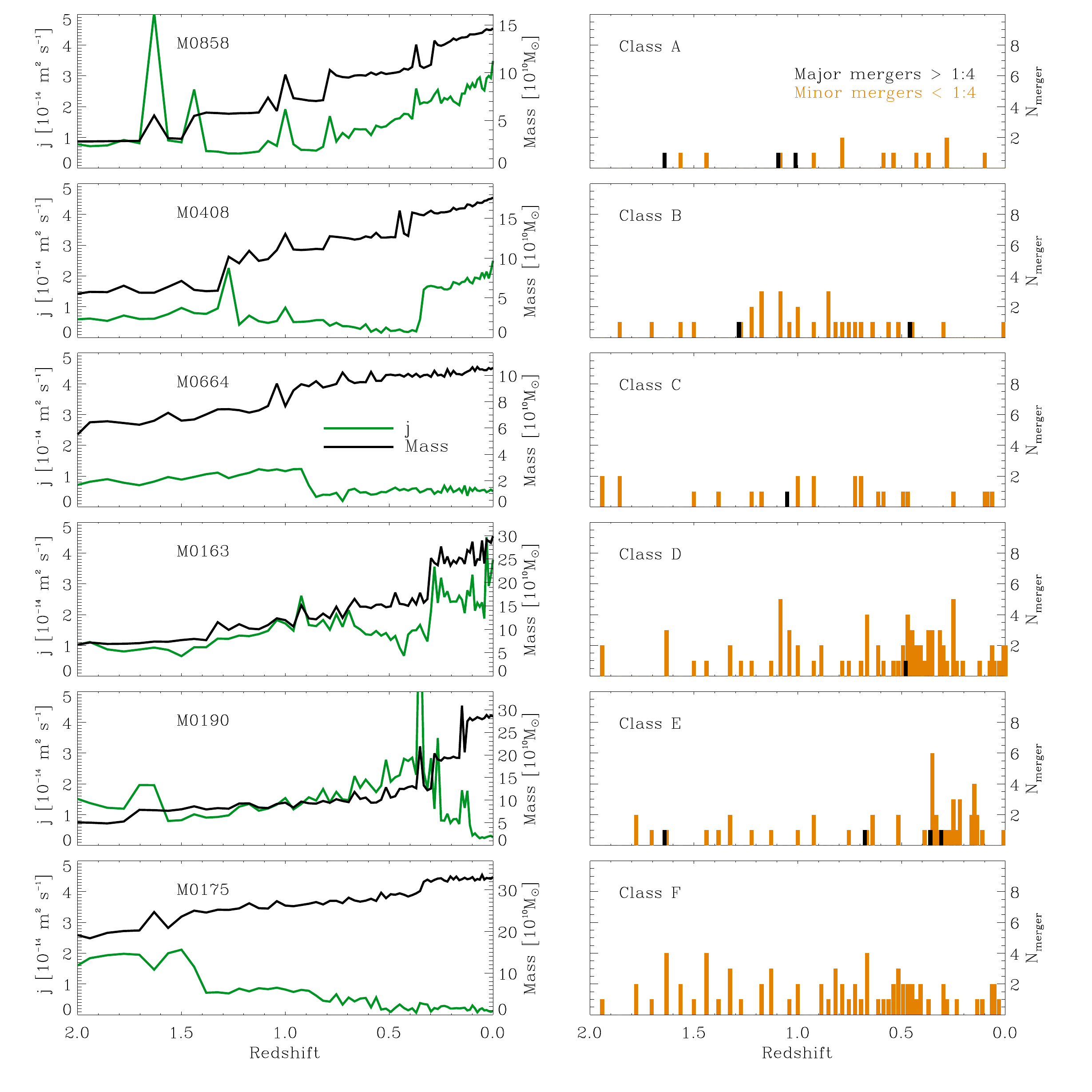, width=0.9\textwidth}
  \caption{ {\it Left panels}: Stellar mass assembly histories (black lines) and
    evolution of the specific angular momenta of the stellar
    components of the galaxies
    (green lines) M0858, M0408, M0664, 
    M0163, M0190, and M0175, from top to bottom. {\it Right panels}:
    Galaxy merger histories represented 
    by the number $N_{\mathrm{merger}}$ of major (black, mass-ratio $\ge$
    1:4) and minor (orange, mass-ratio $<$ 1:4) stellar mergers,
    as a function of redshift for the same galaxies as in the
    respective left panels. Most galaxies experience major mergers
    since $z \approx 2$, only galaxies of class F grow by minor
    mergers alone (M0175, bottom panels, the halo merger tree of this galaxy
    is shown in Fig. \ref{TreeVis}). 
  }
\label{all_ang_merg_evol}
\end{center}
\end{figure*}

\section{Linking assembly histories to present day shapes and
  kinematics}  
\label{linking}

\begin{figure}
\begin{center}
  \epsfig{file=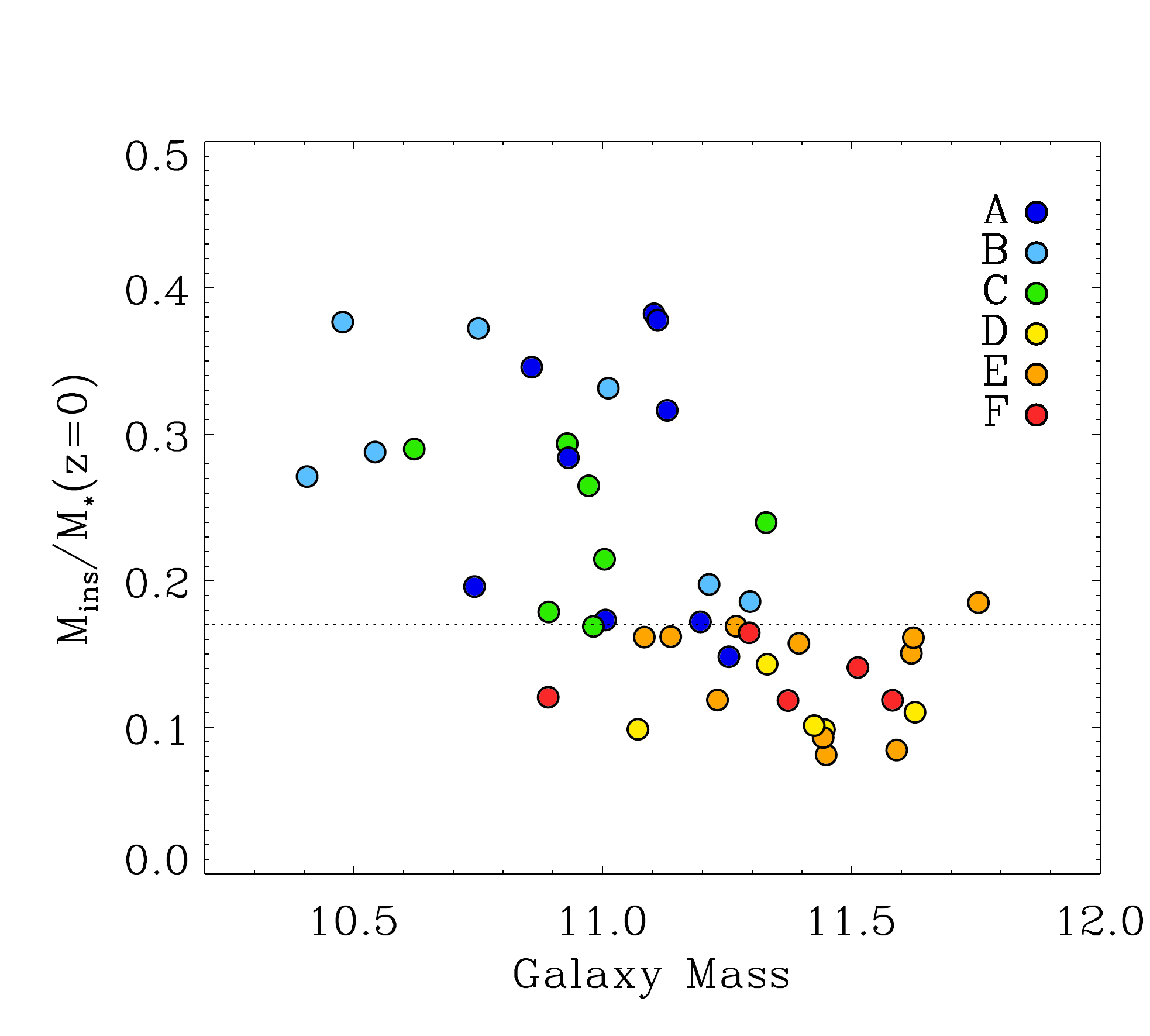, width=0.5\textwidth}
  \caption{Fraction of stars formed in-situ, $M_{\mathrm{ins}}$, since
    $z=2$ to the total present day stellar mass $M_*(z=0)$ for all
    central galaxies sorted by their assembly classes. Galaxies with
    in-situ fractions higher than a fiducial value of 18 per cent
    (horizontal dotted line) show distinct features of dissipative
    star formation in their present day kinematic maps (classes A, B,
    and C, see Fig. \ref{gallery-extra} and Figs. \ref{maps0} -
    \ref{maps7}). The assembly of galaxies with lower in-situ
    fractions (classes D, E, and F) is dominated by accretion and
    merging of stellar systems. These systems are also more massive.}
\label{gm-insfrac-z2}
\end{center}
\end{figure}

\begin{figure}
\begin{center}
  \epsfig{file=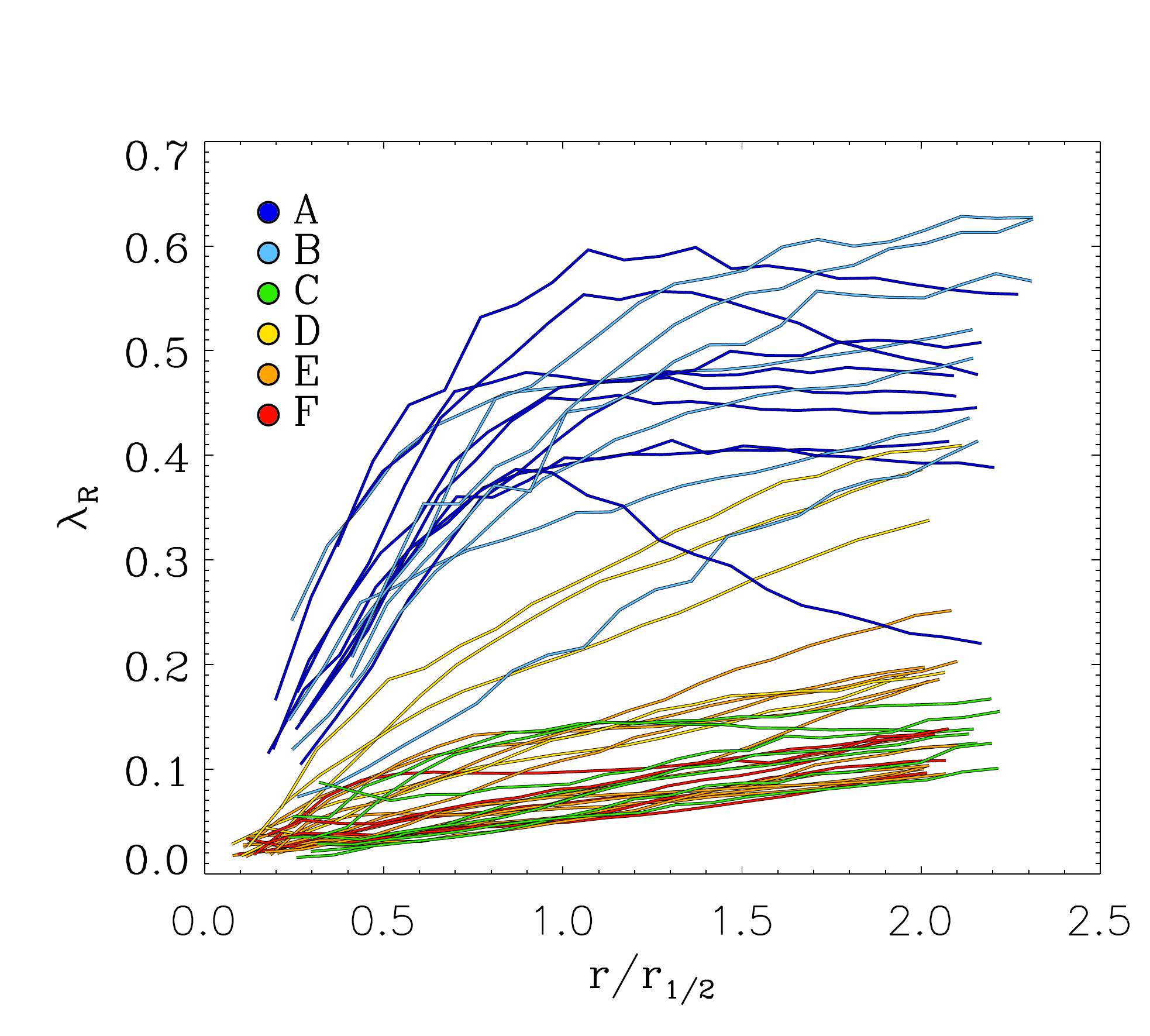, width=0.5\textwidth}
  \caption{$\lambda_{\mathrm{R}}$ profiles for the simulated central galaxies
    up to two half-mass radii sorted by their assembly class. The
    profiles of class A (dark blue) peak at radii $ \lesssim
    r_{\mathrm{e}}$ with no further increase 
    whereas class B profiles (light blue) continuously increase, similar
    to class D. The slow rotators (classes C: green, E: orange F: red)
    have flat or slowly rising profiles. The amplitudes of
    $\lambda_{\mathrm{R}}$ as well as the characteristic profile
    shapes are in agreement with observed early-type galaxies \citep{2004MNRAS.352..721E,2011MNRAS.414..888E}. }  
\label{lambda_profiles_ludwig}
\end{center}
\end{figure}

We have grouped all central galaxies into six classes with distinct
evolutionary paths which can be directly linked to their present day
shapes and kinematics. For this classification we have used the
two-dimensional kinematic maps - one typical example for each class is
shown in  Fig. \ref{gallery-extra} - in combination with the corresponding histories
of mass assembly, specific angular momentum, and galaxy mergers which
are shown in Fig. \ref{all_ang_merg_evol}. The fraction of the total
stellar mass at $z=0$ that formed in-situ since $z=2$ has emerged as a
good measure for the importance of 
dissipation during the late assembly of the galaxies
(Fig. \ref{gm-insfrac-z2}). Galaxies with an in-situ fraction higher
than a fiducial value of 18 per cent show distinct kinematic
features in the kinematic maps. In our simulations the assembly of
galaxies with lower in-situ fractions is dominated by stellar mergers
and dissipationless 
accretion events with clear signatures of these processes in their
present day properties consistent with the trend for the overall
population seen in \citet{2011MNRAS.417..845K}. The radial profiles of the
$\lambda_{\mathrm{R}}$-parameter and the global values inside one
effective radius (as shown in Figs. \ref{lambda_profiles_ludwig} and 
\ref{eps-lambdar}) reveal details about the angular momentum
distribution of the stars. In addition, we analyse for all galaxies the
correlation between the individual pixel values of $v/\sigma$ and
$h_3$ inside one effective radius, which is a measure of the local
asymmetry in the line-of-sight velocity distribution
(Fig. \ref{voversigma_h3_all_3panels}). In the following we first
present the global properties of these six classes (summarised for
each galaxy in Table \ref{tab1}) followed by a more detailed analysis. 

\begin{itemize} 

\item {\bf Class A: Fast rotators with gas-rich minor mergers and
  gradual dissipation.}  
  These galaxies have late ($z < 2$) assembly histories which are 
  dominated by minor - and occasionally early (z $\gtrsim$ 1) major
  mergers (see the example galaxy M0858 in
  Fig. \ref{all_ang_merg_evol}) and a significant amount (up to
  $\sim 40$ per cent) of central in-situ, 
  dissipative, star formation (dark blue dots in
  Fig. \ref{gm-insfrac-z2}). All these galaxies are regular fast 
  rotators with  $0.26 \lesssim \lambda_{\mathrm{R}} \lesssim 0.6$ and
  edge-on ellipticities $0.3 \lesssim \epsilon \lesssim 0.55$
  (Fig. \ref{eps-lambdar}). They are special in showing peaked
  $\lambda_{\mathrm{R}}$ profiles (Fig. \ref{lambda_profiles_ludwig})
  resulting from fast rotating  central disc-like stellar
  configurations. A typical example (M0858) 
  is shown in the first row of Fig. \ref{gallery-extra}. The
  line-of-sight velocity is enhanced towards the center coinciding
  with a clear kinematic signature of a dynamically cold disc, i.e. a
  depression of the stellar line-of-sight velocity dispersion along
  the major axis. The enhanced dispersion above and below the disc
  plane results in a  characteristic dumbbell feature which is
  typical for this class. This galaxy class also shows the most
  asymmetric line-of-sight velocity profiles with a clear
  anti-correlation (steep leading wings) of the local pixel values of
  $v/\sigma$ and $h_3$ (blue dots   in
  Fig. \ref{voversigma_h3_all_3panels}). Additional members of this
  group are M0259, M0380, M0549, M0721, M0908, M1192, M1306, and M1646 
  (see section \ref{samplemaps} for all kinematic maps).    

\item {\bf Class B: Fast rotators with late gas-rich major mergers.}
  Similar to class A the  assembly of galaxies in class B has
  involved significant in-situ star formation with fractions $\gtrsim 0.18$
  (light blue dots in Fig. \ref{gm-insfrac-z2}). In general, these galaxies have
  experienced a late gas-rich major merger leading to a net spin-up of
  the merger remnant or leaving a previously rapidly rotating system
  unchanged (see e.g. M0408, second row of
  Fig. \ref{all_ang_merg_evol}). The $\lambda_{\mathrm{R}}$ values and
  ellipticities  (light blue dots in Fig. \ref{eps-lambdar}) are
  in the same range as those of class A but the $\lambda_{\mathrm{R}}$
  profiles (light blue lines in Fig. \ref{lambda_profiles_ludwig}) are 
  constantly rising beyond $r_e$. Galaxies of class A and class B have
  the youngest mass-weighted
  stellar populations ($\sim 9.5$ Gyrs) of the whole sample. Similar to
  class A, $h_3$ and $v/\sigma$ are anti-correlated  (light blue dots
  in the left panel of Fig. \ref{voversigma_h3_all_3panels}). All
  galaxies in this class have additional signatures, but not as strong as
  class A, of embedded stellar disc components  including disc-like
  velocity fields, mid-plane depressions in the stellar velocity
  dispersions, and, occasionally, pointy isophotes. Additional members
  of this class are M0290, M0977, M1196, M3852, M5014, and M6782.  

\item {\bf Class C: Slow-rotators with late gas-rich major mergers.} This class contains
  all galaxies that have experienced a late gas-rich major merger
  leading to a spin-down of the stellar remnant or leaving the spin 
  of a slowly rotating progenitor unchanged. A typical example is
  M0664 (third row of Fig. \ref{gallery-extra} and
  Fig. \ref{all_ang_merg_evol}). Class C galaxies also have high
  in-situ fractions  (similar to classes A and B, green dots in
  Fig. \ref{gm-insfrac-z2}) with typical central depressions in the
  stellar velocity dispersion. This feature originates from stars that
  have formed from gas driven to the 
  center  of the galaxy during the merger, a process well studied in
  isolated binary mergers \citep{1996ApJ...471..115B}. The galaxies
  rotate slowly (green lines in Fig. \ref{lambda_profiles_ludwig}) and
  are among the roundest in our sample with edge-on ellipticities of
  $\epsilon \sim 0.3$ (green dots in Fig. \ref{eps-lambdar}), again an  
  effect of gas  dissipation
  \citep{1996ApJ...471..115B,2006ApJ...641...21R}. Additional galaxies
  in this group are M0443, M0959, M1017, M1071, M1167, M2665.

\item {\bf Class D: Fast-rotators with late gas-poor major mergers.} All
  galaxies in this class have, in addition to minor mergers,
  experienced a recent collisionless major merger leading to a
  significant spin-up of the stellar remnant or leaving the properties
  of a previously fast rotating galaxy unchanged ($0.1 \lesssim  \lambda_{\mathrm{R}}
  \lesssim  0.3$). An example (M0163) is shown in the fourth row of
  Figs. \ref{gallery-extra} and \ref{all_ang_merg_evol}. In contrast to 
  galaxies of class B the late assembly and the merger did not involve 
  significant amounts of gas resulting in a low global in-situ
  fraction (yellow dots in Fig. \ref{gm-insfrac-z2}). Despite their
  fast rotation, galaxies in this class show no additional signatures
  for embedded disc-like components and the LOSVDs do not have steep
  leading wings (no clear anti-correlation of $v_{\mathrm{los}}$ and
  $h_3$ in the second panel of
  Fig. \ref{voversigma_h3_all_3panels}). This particular feature has
  been investigated in detail for binary mergers and is
  characteristic for fast rotating remnants of mergers without gas 
  \citep{2001ApJ...555L..91N,2006ApJ...636L..81N,2006MNRAS.372..839N,2007MNRAS.376..997J}. 
  Additional galaxies in this class are M0040, M0163, M0224, M0227,
  M0763.

\item {\bf Class E: Elongated slow-rotators with late gas-poor major mergers.}
  Galaxies in this class have, in addition to minor mergers,
  undergone at least one recent recent major merger which has
  lead to a significant spin-down of the remnant or has only mildly changed the 
  properties of a previously slowly rotating galaxy (e.g. M0190, fifth
  row in Figs. \ref{gallery-extra} and \ref{all_ang_merg_evol}). Their late 
  assembly involved little dissipation (orange dots in
  Fig. \ref{gm-insfrac-z2}) and all galaxies in this class are slowly 
  rotating  $\lambda_{\mathrm{R}} \lesssim  0.19$ with slowly rising
  $\lambda_{\mathrm{R}}$-profiles (orange lines in
  Fig. \ref{lambda_profiles_ludwig}). Ellipticities are significantly
  higher than for galaxies in class C (which have similar merger
  histories but more dissipation) with $0.3 \lesssim \epsilon \lesssim 0.5$
  (orange dots in Fig. \ref{eps-lambdar}). The properties are
  consistent with results from binary collisionless major merger
  simulations with remnants that are slowly rotating but have a
  prolate shape and, occasionally, show strong kinematic twists
  (M0215) minor-axis rotation like M0190 \citep{1992ApJ...400..460H,2006ApJ...650..791C}.  
  Other galaxies in this class are
  M0053, M0069, M0089, M0162, M0215, M0300, M0501, M0616, M0094, M0209. The 
  only clear counter-rotating core in our sample is a galaxy of this
  class (M0094).  

\item {\bf Class F: Round slow-rotators with gas-poor minor mergers only.} The $z
  \lesssim 2$ assembly  history of these galaxies is dominated by
  stellar minor mergers without any major mergers (e.g. M0175, see
  bottom row of Fig. \ref{all_ang_merg_evol}) and little in-situ star
  formation (red dots in Fig. \ref{gm-insfrac-z2}). Galaxies of this
  class have the lowest angular momentum $\lambda_{\mathrm{R}}
  \lesssim 0.09$ with almost featureless velocity fields (bottom row
  of Fig \ref{gallery-extra}) and are among  the roundest galaxies in our sample
  with $\epsilon \sim 0.27$ (red dots in Fig. \ref{gm-insfrac-z2}). We
  find four more galaxies with similar properties in our sample 
  (M0125, M0204, M0329, M0984).       

\end{itemize}

\begin{deluxetable}{ccccccccccc}
\tablewidth{0pt}
\tablecaption{Properties of galaxy classes \label{tab:classes}}
\tablehead{
 \colhead{Class} &
 \colhead{$M_*$} & 
 \colhead{$\lambda_{\mathrm{R}}$} &
 \colhead{$\lambda_{\mathrm{R}}$-profile} &
 \colhead{$\epsilon$} &
 \colhead{$M_{ins}/M_{*}$} &
 \colhead{$\langle age \rangle$} &
 \colhead{Mergers} & 
 \colhead{$h_3$ - $v/\sigma$} &
 \colhead{map-features} & \\
 \colhead{}  \\
 \colhead{(1)} &
 \colhead{(2)} &
 \colhead{(3)} &
 \colhead{(4)} &
 \colhead{(5)} &
 \colhead{(6)} &
 \colhead{(7)} &
 \colhead{(8)} &
 \colhead{(9)} 
}
\startdata
A  & 11 & 0.45 & peaked    & 0.45 &  0.27 & 9.7 & mj \& mi & strong & dumbbell \\
B  & 8.7 & 0.47 & rising   & 0.45 &  0.29 & 9.3 & mj \& mi & strong  & discs \\
C  & 10 & 0.10 & flat      & 0.31 &  0.24 & 9.9 & mj \& mi &  no  &
dispersion dip  \\
D  & 26 & 0.23 & rising    & 0.36 &  0.11 & 10.6 & mj \& mi &  very
weak & fast rotation \\
E  & 29 & 0.11 & slowly rising  & 0.43 &  0.14 & 10.7& mj \& mi &
no & slow rotation \\
F  & 24 & 0.08 & flat      & 0.27 & 0.13 & 10.9 & mi only & no & no rotation \\
\enddata
\tablecomments{
(1): Assembly class as discussed in the text.
(2): Mean stellar mass inside $R_{10}$ in $10^{10} M_{\odot}$.
(3): Mean value of $\lambda_{\mathrm{R}}$.
(4): Shape of the $\lambda_{\mathrm{R}}$-profile. 
(5): Mean ellipticity. 
(6): Mean in-situ mass fraction; dissipative assembly
for galaxies with a fiducial value larger than 18 per cent.   
(7): Mean mass-weighted stellar age in Gyrs.  
(8): Mergers relevant for galaxy assembly; mj: major mergers, mi:
minor mergers. 
(9): Strength of the anti-correlation between $h_3$  and
$v/\sigma$.
(10): Special features in the kinematic maps.
}
\end{deluxetable}
\clearpage

We have summarized the global properties of these classes in
Table \ref{tab:classes} and now present the detailed analysis on which 
this classification is based. In Fig. \ref{gallery-extra} we
show characteristic examples of the two-dimensional maps for the line-of-sight velocity
($v$), line-of-sight velocity dispersion ($\sigma$), a measure for the
asymmetric ($h_3$), and symmetric ($h_4$) deviations from a Gaussian
LOSVD shape, within two $R_{1/2}$ for one galaxy of each class described
above (M0858, M0408, M0664, M0163, M0190, M0175) from top to bottom,
see Table \ref{tab1} for individual galaxy properties). The kinematic
maps for the central galaxies of the whole sample (including the
galaxies shown here) are presented in Figs. \ref{maps0} - \ref{maps7}
of Section \ref{samplemaps}. For these maps rotation, kinematic
substructure, and higher-order kinematic features are clearly visible,
quantifiable and have similar amplitudes as in observed real galaxies.   

M0858 (top row in Fig. \ref{gallery-extra}) is an example of a fast
rotator of class A. The velocity field shows peaked regular rotation
of a disc-like component, also visible as a depression
(dumbbell-feature) of the stellar velocity dispersion along the major
axis. The LOSVDs along the major axis show steep leading wings with a
clear signature of an anti-correlation between $v_{\mathrm{los}}$ and
$h_3$ and $h_4$ is predominantly positive in the disc region. M0408
(second row in Fig. \ref{gallery-extra}) is also a fast rotator (class
B). In contrast to M0858 rotation here extends to larger radii and
the depressions in the velocity dispersion are only evident at $ >
r_{\mathrm{e}}$. This disc-feature is accompanied by pointy, discy,
isophotes and anti-correlated $h_3$. A galaxy with only weak rotation
in the inner part is M0646 (class C) with a central depression in the
velocity dispersion, no features in $h_3$, and positive central values for
$h_4$. M0163 (class D) is classified as a fast rotator but does not
show disc-like rotation similar to M0858 or M0408. The dispersion profile
is peaked and there are no (weak) features in the $h_3$ and $h_4$ maps. A
very elongated but non disc-like and slowly rotating system is M0190
(class E) which also has a centrally peaked velocity dispersion and no
features in $h_3$ and $h_4$. One of the roundest, most slowly
rotating, and most featureless galaxies in the sample is M0175 (class
F, bottom row of Fig. \ref{gallery-extra}).   

In the left panels of Fig. \ref{all_ang_merg_evol} we show the stellar
mass assembly histories (black) and the evolution of the specific angular
momentum of all stars within the effective radius (green) (which
correlates well with $\lambda_{\mathrm{R}}$
i.e. \citealp{2009MNRAS.397.1202J}) since $z \sim 2$. Fluctuations in
the values can be attributed to ongoing mergers and interactions. The
corresponding stellar merger histories (black: major mergers with
mass-ratios $> 1:4$, orange: minor mergers with mass-ratios $<$ 1:4)
are presented in the right panels. Every merger is caused by an
increase of stellar mass. This is most obvious for major
mergers but also visible for minor mergers. However, the angular
momentum can both decrease (M0664, M0190) or increase during a major
merger (M0408, M0163). The most extreme examples are M0190 and M0408. For 
the former two major mergers at $z \sim 0.3$ cancel all rotation in
a previously rapidly rotating galaxy with a concurrent mass increase
by a factor of $\sim 2.5$. The latter is almost non-rotating before
the merger at $z \sim 0.4$ and spins up significantly. In some cases 
the mass increase is not accompanied by mergers and originates from
gas accretion and in-situ star formation which in general is
accompanied by an increase of the stellar angular momentum
(e.g. M0858). For some massive galaxies the mass assembly history is
dominated by minor mergers. The example presented here is M0175
(bottom panels of Fig. \ref{all_ang_merg_evol}). This galaxy has no
major mergers and the mass increase by a factor of $\sim 1.8$ since $z
\sim 2$ is solely driven by minor mergers continuously transforming a
fast rotating galaxy into a slow rotator at the present day.     

\begin{figure}
\begin{center}
  \epsfig{file=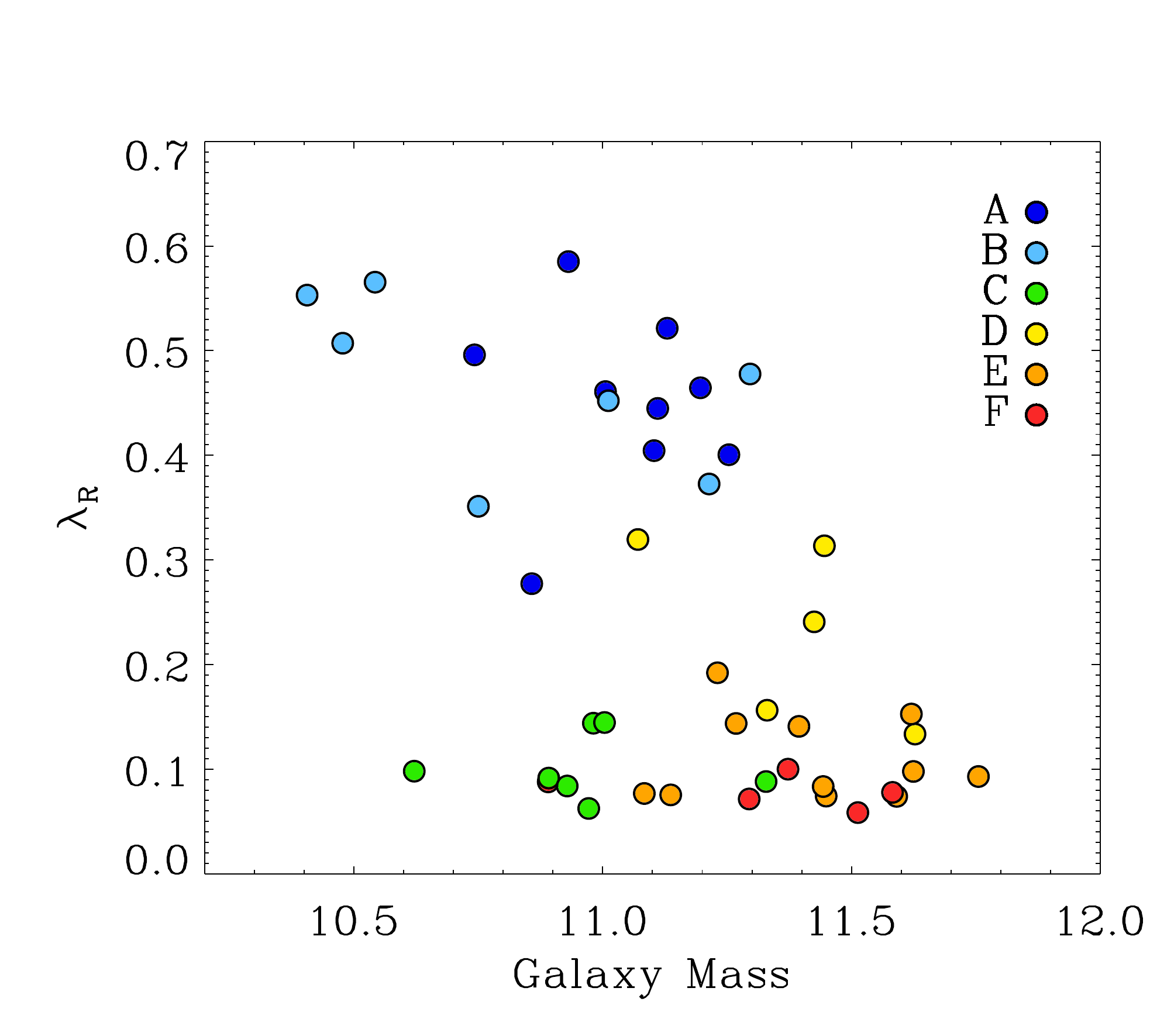, width=0.5\textwidth}
  \caption{Global edge-on $\lambda_{\mathrm{R}}$-parameter versus stellar
    galaxy mass. As in Fig. \ref{gm-insfrac-z2} the galaxies are color
    coded by their assembly class. Galaxies involving dissipation in
    their late formation (class A, B, and C) dominate lower
    masses. Galaxies with late dissipationless assembly (classes D, E,
    and F) form the most massive systems with a lower specific angular
    momentum.}
\label{gm-lambdar}
\end{center}
\end{figure}

\begin{figure}
\begin{center}
  \epsfig{file=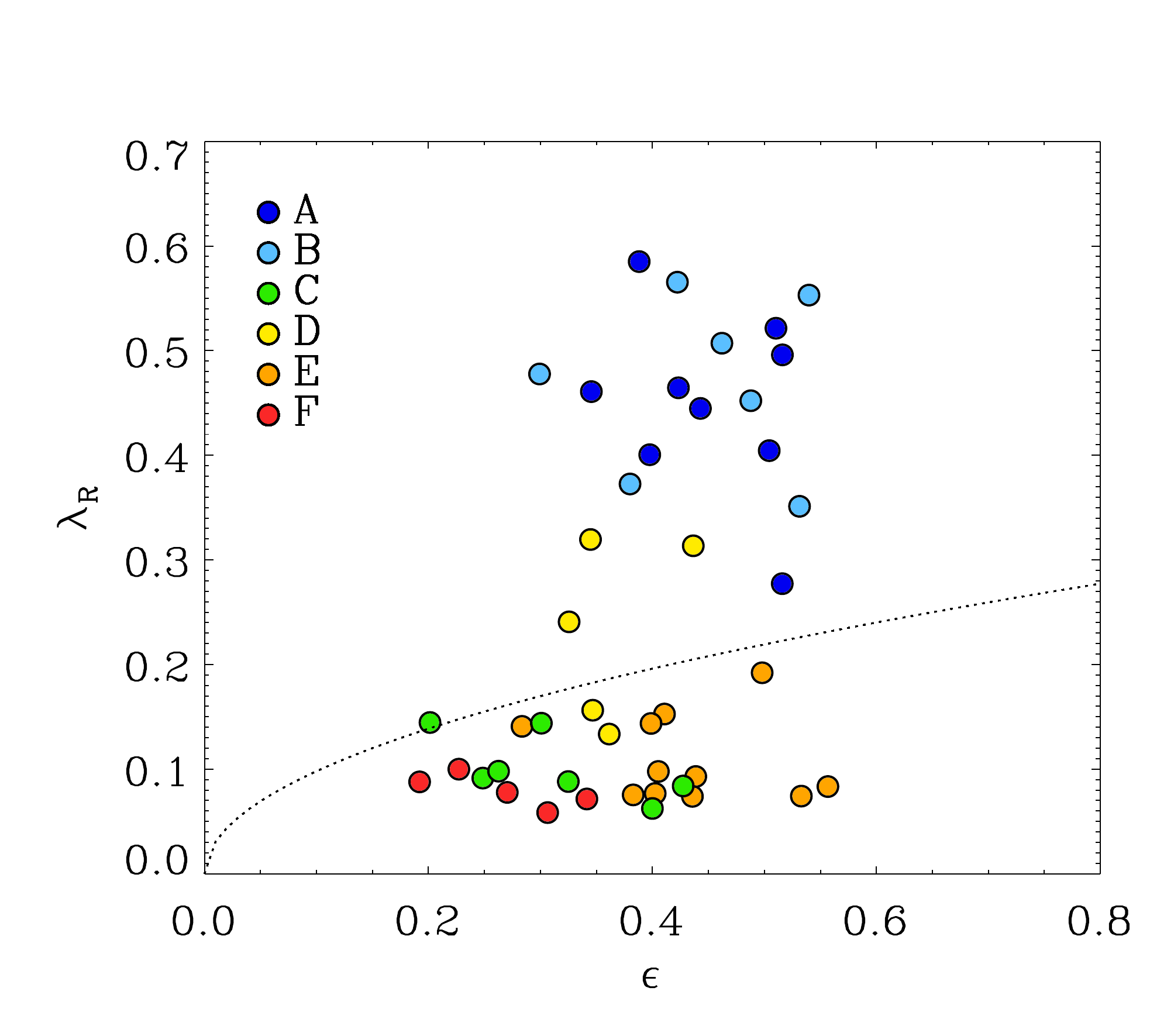, width=0.5\textwidth}
  \caption{The $\lambda_{\mathrm{R}}$-parameter versus projected
    edge-on ellipticities at $r_{\mathrm{e}}$. As in
    Fig. \ref{gm-insfrac-z2} the galaxies are color coded by their
    assembly class. Galaxies of class A and B (blue colors) are all
    elongated fast rotators. Galaxies of class D are rounder and
    rotate more slowly. The most elongated slow rotators are class E
    galaxies (orange). Galaxies of class C (green) show similarly low
    rotation but are rounder. The slowest rotators and roundest
    galaxies are those of class F (red), whose assembly history is
    dominated by minor mergers alone.}
\label{eps-lambdar}
\end{center}
\end{figure}

To estimate the amount of dissipation involved in the assembly of the
stellar bodies of the galaxies we identify the stellar mass
formed in-situ, $M_{\mathrm{ins}}$, in the galaxy since $z \sim 2$ (see
\citealp{2010ApJ...725.2312O}). The fraction of this mass to the total
present day stellar mass $M_*(z=0)$ 
for all central galaxies (colored by their assembly class) versus the
present day stellar mass is shown in Fig. \ref{gm-insfrac-z2}. In
general, more massive galaxies have fewer in-situ assembled stars
\citep{2010ApJ...725.2312O} and for most galaxies of classes D, E, and
F  this value is below a fiducial value of 18 per cent. Their assembly
is dominated by accretion and merging of stellar systems. Galaxies of
classes A, B, and C typically have lower masses and have more stars
formed in-situ since $z \sim 2$. They also show distinct features of
dissipative disc-like star formation in their present day kinematic maps (classes A, B, and
C, see Fig. \ref{gallery-extra} and Fig. \ref{maps0} - \ref{maps7}). 


\begin{figure}
\begin{center}
  \epsfig{file=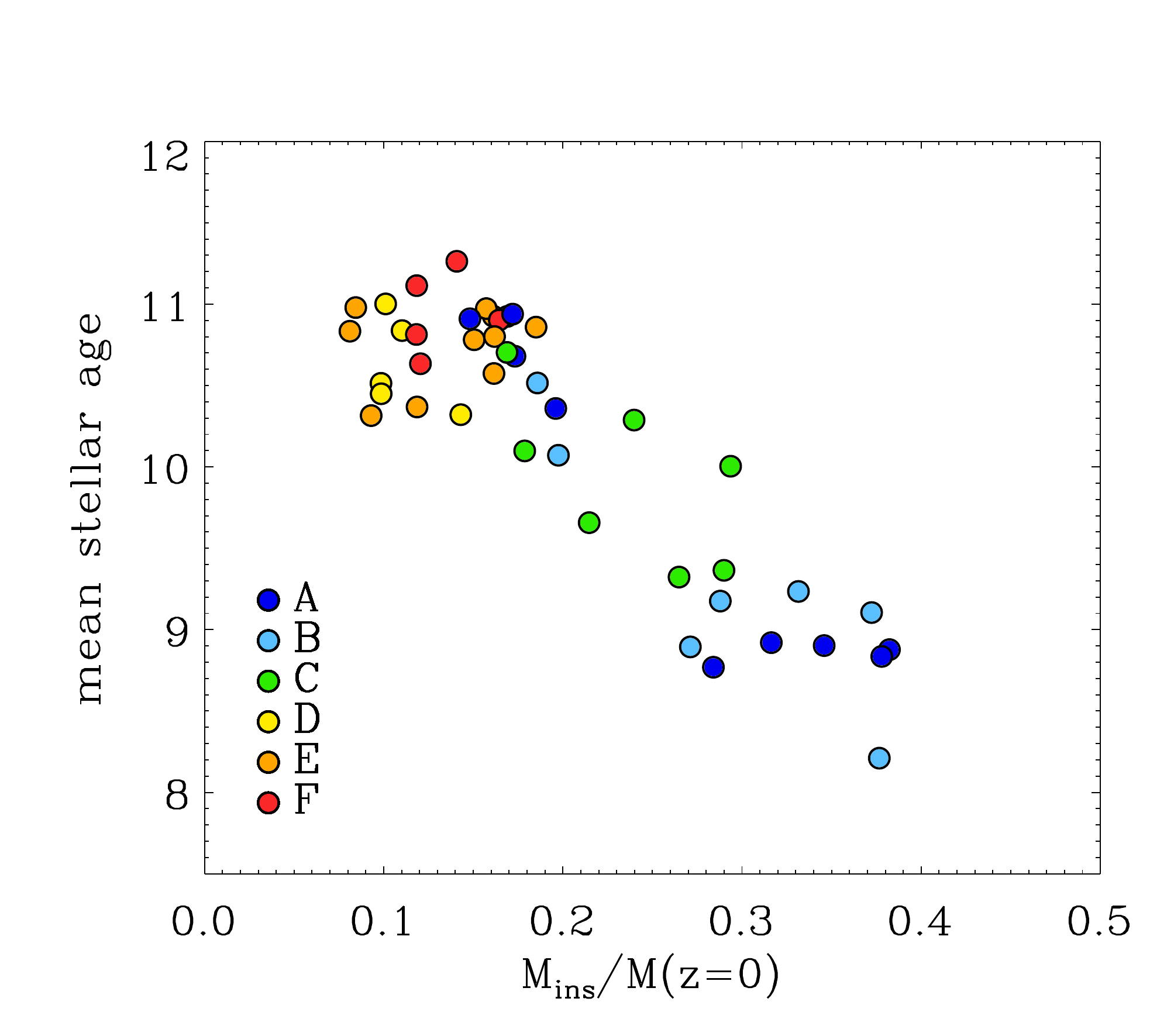, width=0.5\textwidth}
  \caption{Mass-weighted stellar ages of the central 
    galaxies versus the fraction of stars formed in-situ,
    $M_{\mathrm{ins}}$, since $z=2$. Galaxies of assembly classes D,
    E, and F with predominantly dissipationless recent assembly
    histories (see Fig. \ref{gm-insfrac-z2}) are consistently old
    ($\sim 10.7$ Gyrs). As expected galaxies of classes A, B, and C
    (whose late assembly involves more dissipation, see
    Fig. \ref{gm-insfrac-z2}) in general are younger and show a larger
    spread in age (some have ages similar to classes D, E, and F, some
    are as young as $\sim 8.5$ Gyrs).}
\label{ins-meanage}
\end{center}
\end{figure}

%
%

\begin{figure*}
\begin{center}
  \epsfig{file=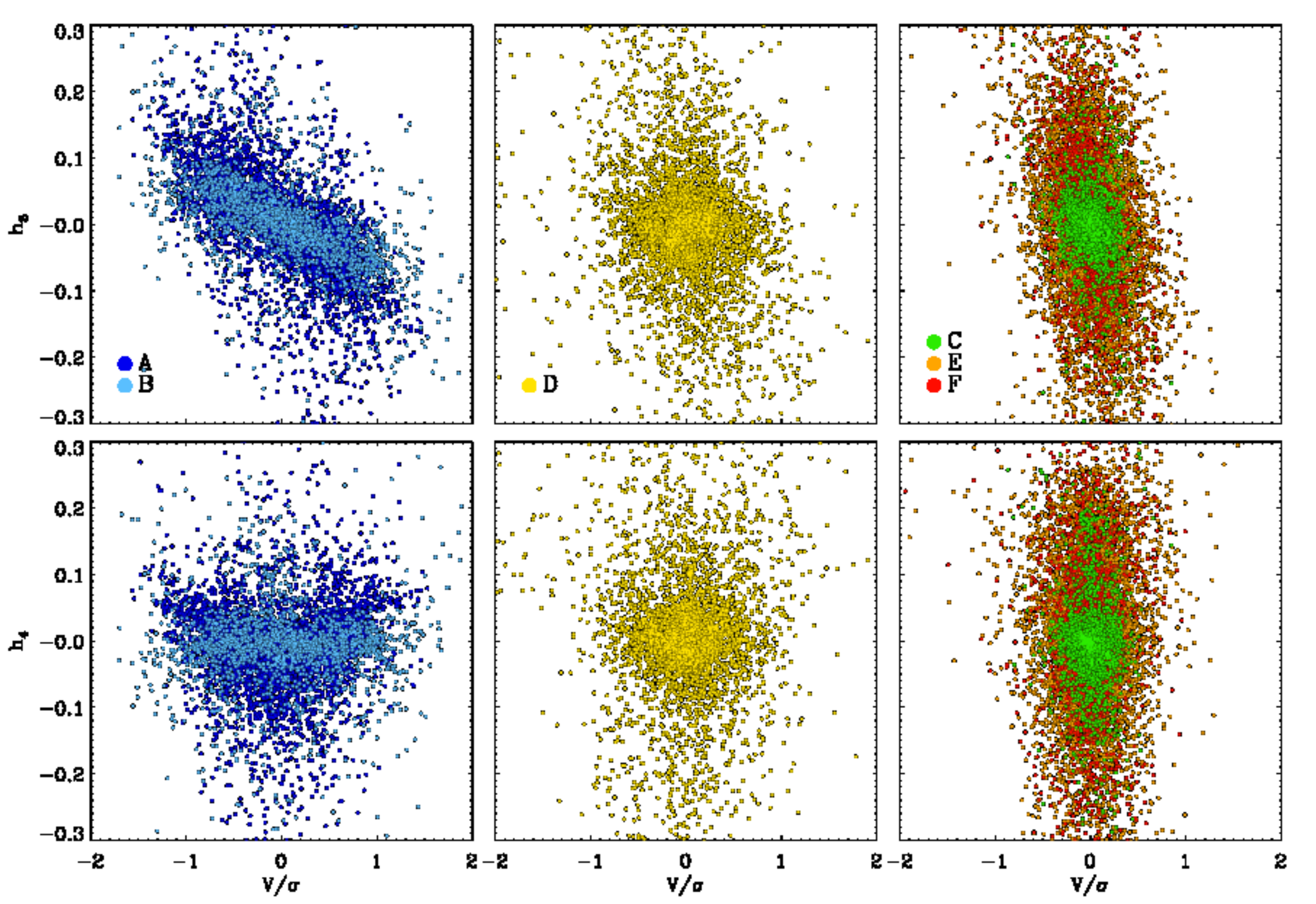, width=0.9\textwidth}
  \caption{Local pixel values within $r_{\mathrm{e}}$ of $h_3$ (upper
    panels) and $h_4$ (lower panels) versus $v/\sigma$ for all
    galaxies of assembly classes A and B (left 
    panels), class D (middle panels, and classes E, C, and F (right
    panels). Maps of fast rotating class A/B galaxies clearly
    show an anti-correlation indicating steep leading wings in the
    local line-of-sight velocity distributions caused by axisymmetry
    and/or the presence of a disc-like component. The amplitude of
    $h_4$ is low and class A shows a V-shaped structure. Galaxies of class D
    can also be rotation dominated but show no correlation between
    $h_3$ and $v/\sigma$  - a signature of the recent gas poor major
    merger. Galaxies of classes E, C, and F rotate slowly and
    show no correlations. The amplitude of $h_4$ is in general higher
    than for classes A, B, and C. These results are in good
    qualitative agreement with observed early type galaxies
    \citep{2011MNRAS.414.2923K}.}   
\label{voversigma_h3_all_3panels}
\end{center}
\end{figure*}

These features are also apparent in the variation of
$\lambda_{\mathrm{R}}$ as a function of radius for the edge-on
projections of all simulated central galaxies up to two 
projected half-mass radii (Fig. \ref{lambda_profiles_ludwig}). Galaxies of
class A and B have steeply rising profiles with peaks inside of one
$r_{\mathrm{e}}$ for class A. Galaxies of class C (green) have lost their
angular momentum in a gas-rich merger and have slowly rising profiles,
similar to slowly rotating galaxies with purely collisionless histories
(classes E and F). The exception are galaxies of class D that have
gained angular momentum in a recent major merger. The amplitude of 
$\lambda_{\mathrm{R}}$ as well as the characteristic profile shapes
are in good qualitative agreement with observed early-type galaxies
\citep{2004MNRAS.352..721E,2011MNRAS.414..888E}. Based on the same
simulations \citet{2012arXiv1209.3741W} find that the trends in
$\lambda_{\mathrm{R}}$ extend to even larger radii.

The distribution of galaxies in the $\lambda_{\mathrm{R}}$-$M_*$ plane
in shown in Fig. \ref{gm-lambdar} for the galaxies seen edge-on
(projection effects are discussed in Section \ref{projection}). 
Galaxies with the
highest stellar masses are all slowly rotating and mostly belong to
classes E and F.  At lower stellar masses the distribution is bi-modal
with a group of galaxies with slow rotation  $\lambda_{\mathrm{R}}
\lesssim  0.15$  all belonging to group C and 
another group of galaxies (mostly from group A and B) showing rapid
rotation $\lambda_{\mathrm{R}} \gtrsim 0.25$. These results indicate
that slow rotators at low stellar masses have predominantly formed
from dissipative major mergers whereas at high stellar masses the
recent assembly history was essentially collisionless. Similar
conclusions have been reached by \citet{2011MNRAS.417..845K} looking 
at the fast and slow rotator demographics in semi-analytical models.     

In Fig. \ref{eps-lambdar} we show the location of the central
galaxies in the $\lambda_{\mathrm{R}}$ - $\epsilon$ plane color coded
by their assembly classes. Here the characteristic values of
$\lambda_{\mathrm{R}}$ as well as $\epsilon$ have been determined at
one effective radius. Galaxies with dissipative histories (classes A
and B) are the fastest rotators with ellipticities $0.3 \lesssim
\epsilon \lesssim 0.58$. Galaxies of class D have intermediate
properties. The slowest rotators are classes C (gas-rich major
merger), E (gas-poor major merger) and F (only minor
mergers). The roundest ($\epsilon \sim  0.27$) and most
slowly rotating ($\lambda_{\mathrm{R}} \sim 0.08$) galaxies are those of
class F whose $z < 2$ mass assembly is dominated by minor
mergers. Slow rotators with late gas-poor major mergers (class F) are the most
flattened with ellipticities similar to fast rotators.     
Galaxies of this class are also the oldest at an average age of 10.9
Gyrs. In Fig. \ref{ins-meanage} we show the mean age as a function of 
the in-situ mass-fraction color coded by assembly class. The overall
correlation is almost by construction. Still, galaxies of classes D,
E, and F have consistently old ages ($\sim 10.7$ Gyrs) whereas the age
spread for classes A, B, and C is significant (8.2 Gyrs $\lesssim$ age
$\lesssim$ 11.0 Gyrs).   

In addition, we present for all galaxies the
correlation between the individual pixel values of $v/\sigma$ and
$h_3$ as well as $h_4$ inside one effective radius
(Fig. \ref{voversigma_h3_all_3panels}). A strong anti-correlation of
$h_3$ and $v/\sigma$ indicates line-of-sight velocity distributions with
steep leading wings indicative of a rotating  axisymmetric stellar
body and, eventually, a stellar disc (see
e.g. \citealp{1994MNRAS.269..785B}).  A positive (negative) value for $h_4$
indicates a velocity distribution that is more (less) peaked than a
Gaussian. We present the correlations for
three galaxy groups separately. Galaxies of assembly class A and B
show a similarly strong anti-correlation of $v/\sigma$ and $h_3$.
These galaxies are also fast rotators and with respect to the
anti-correlation, the absolute 
values of $h_3$, and $v/\sigma$ they are reminiscent of observed
properties of regular rotators (\citealp{1994MNRAS.269..785B};
\citealp{2008MNRAS.390...93K}; and see Fig. 9 in
\citealp{2011MNRAS.414.2923K}). Galaxies of group D are also fast
rotating but do not show a clear anti-correlation. This can be
understood in terms of their formation history. Their fast rotation
was caused by a recent gas poor major merger which has been shown to have
the ability to spin up galaxies
\citep{2009A&A...501L...9D,2011MNRAS.416.1654B}. However, due to the
absence of a dissipative gas component during the merger, the remnant galaxies
are typically not axisymmetric and neither support the population
of stars on high angular momentum tube orbits nor the re-growth of a
significant disc component. This explains the absence of a strong
anti-correlation
\citep{1996ApJ...471..115B,2000MNRAS.316..315B,2001ApJ...555L..91N,2007MNRAS.376..997J,2006MNRAS.372..839N,2009ApJ...705..920H,2010ApJ...723..818H}. All
slow rotators (classes C, E, and F) show no correlation with $h_3$ in
agreement with long-slit and integral-field observations
\citep{1994MNRAS.269..785B,2008MNRAS.390...93K,2011MNRAS.414.2923K}.

\section{Projection effects and satellite properties}
\label{projection}

So far we have only studied edge-on projections of the simulated
central galaxies. However, both the $\lambda_{R}$ parameter as well as
the ellipticity $\epsilon$ change with varying viewing angles,
i.e. edge-on galaxies with fast rotation and high ellipticities
typically show a decrease in $\lambda_{R}$ as well as $\epsilon$ as
they are projected more face-on 
\citep{2009MNRAS.397.1202J,2010MNRAS.406.2405B,2011MNRAS.414..888E,2011MNRAS.416.1654B}.   

\begin{figure}
\begin{center}
  \epsfig{file=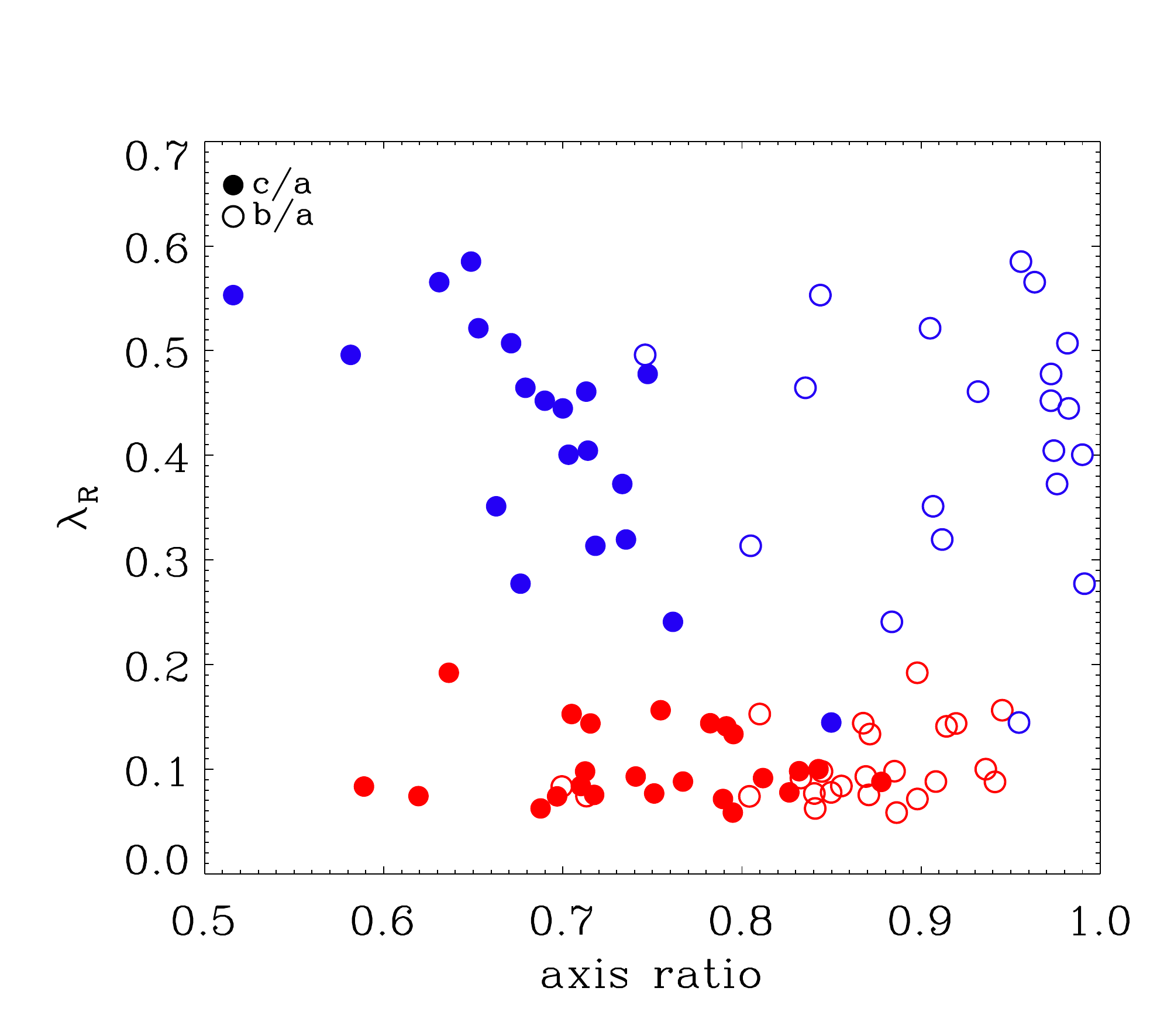, width=0.5\textwidth}
  \caption{The $\lambda_{\mathrm{R}}$-parameter versus the intrinsic shape of the
    inner region (50 \% most tightly bound stars) of the simulated
    central galaxies. Here $c/a$ (filled  
    circles) is the ratio of the short to the long axis and $b/a$
    (open circles) is the ratio of the intermediate to the long
    axis. Fast rotators (blue) are significantly flattened ($c/a < b/a$) and
    nearly oblate with b/a $\gtrsim$ 0.8, whereas slow rotators (blue)
    tend to be more round or slightly triaxial ($c/a \lesssim b/a$). }
\label{shape_lambda_ludwig}
\end{center}
\end{figure}

For an interpretation of projection effects we consider the underlying
three-dimensional shape of the galaxies. We compute the intrinsic
shape of the central stellar component (50\% most tightly 
bound particles) from the principal axes of the moment of inertia
tensor (a: long axis, b: intermediate axis, c: short axis). Fast
rotating galaxies are significantly flattened ($c/a < b/a$) and nearly oblate
with $b/a \gtrsim 0.8$ (Fig. \ref{shape_lambda_ludwig}). With a mean 
flattening $c/a = q \sim 0.6$ the simulated fast 
rotators are significantly rounder than the inferred mean intrinsic
shape of fast rotators in the $ATLAS^{\rm{3D}}$ survey of $q=0.25$ (Weijmans
et al., 2013, submitted). Simulated slow rotators are significantly
more round ($q \sim 0.75$), but still triaxial  ($c/a \lesssim b/a$), when
compared to their fast rotating counterparts. Again, the simulated galaxies
are on average rounder than the observed ($q \sim 0.63$). It is
plausible to assume that the shape difference between model galaxies
and observations originate from the assumed model for star
formation. The employed model favors early conversion of gas into
stars and therefore the formation of bulges. The observed disc-like
properties, in this case the flattened shape, of most local early type
galaxies cannot be recovered accurately.

\begin{figure*}
\begin{center}
  \epsfig{file=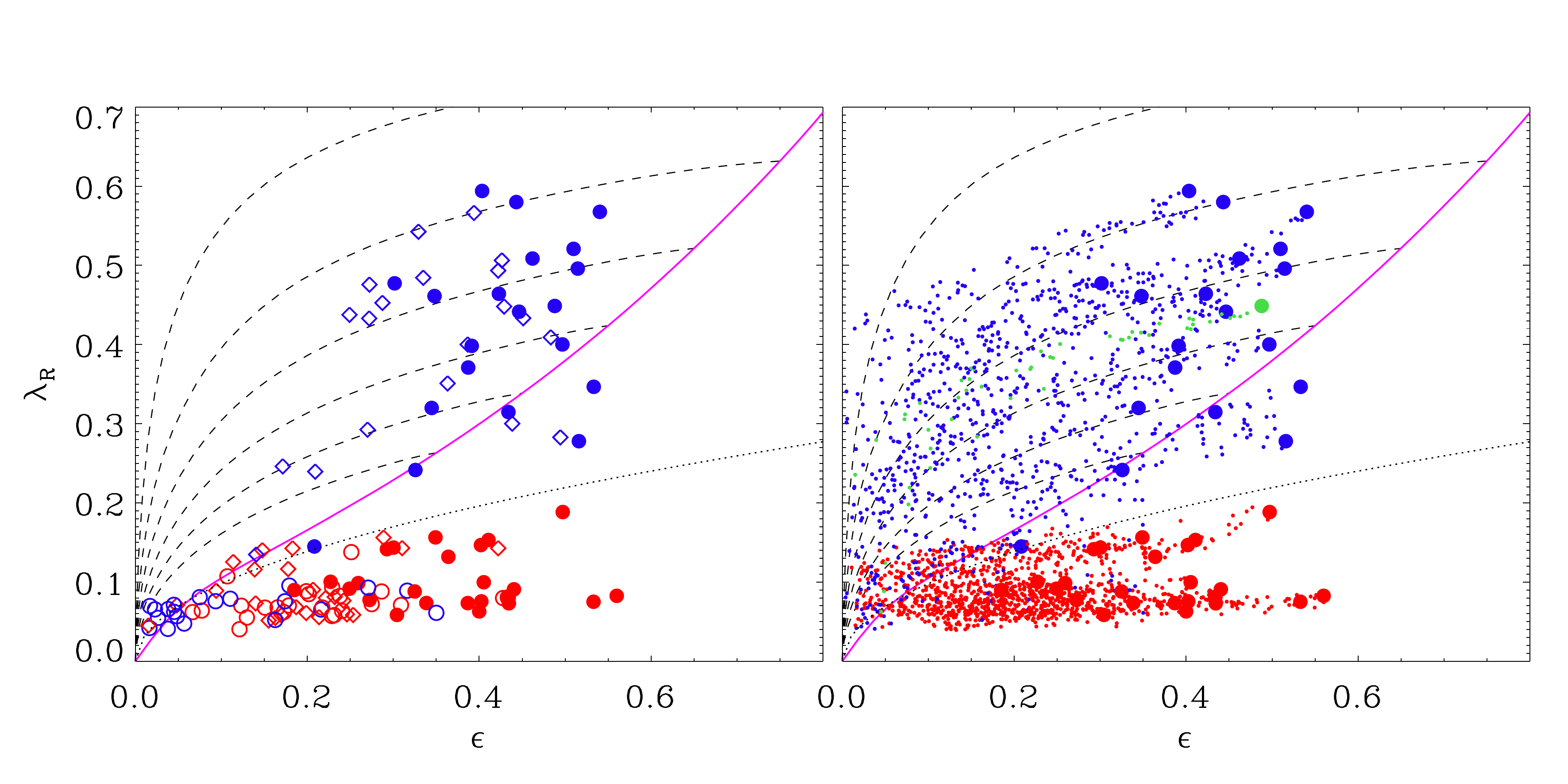, width=1.0 \textwidth}
  \caption{{\it Left:} The $\lambda_{\mathrm{R}}$ parameter
    for fast rotators (blue) and slow rotators (red) as defined from
    the edge-on projection (seen along the intermediate axis) versus
    projected ellipticities of the simulated early-type galaxies. The 
    black dotted division line between fast 
    and slow rotators is based on the empirical relation presented in
    \citet{2011MNRAS.414..888E}. Filled dots indicate the edge-on
    projections, diamonds show the projections along the long axis
    and open circles are face-on projections along the short axis. The 
    magenta solid line \citep{2007MNRAS.379..418C} shows the edge-on
    view integrated up to one effective radius for spheroids with an
    intrinsic anisotropy, $\beta = 0.65 \times \epsilon$, which
    (including projection effects)  encompasses most  observed fast
    rotating spheroids. The black dashed lines indicate the effect of
    projection from edge-on to face-on for ellipticities $\epsilon  =
    0.82, 0.75, 0.65, 0.55, 0.45, 0.35$ from top to bottom. Most of
    our simulated fast rotators cover the same region. {\it Right:}
    Similar to the left panel but for 50 random 
    projections of fast (blue) and slow (red) rotators. Oblate fast rotators project
    towards lower ellipticities in agreement with the analytic
    expectations (black dashed lines). The effect of projection in
    this plane is highlighted for one fast rotator (green dots). Slow
    rotators (red dots) do not change their rotation properties and
    just project towards rounder shapes.} 
\label{ell_lambda_ludwig_good}
\end{center}
\end{figure*}

The difference in intrinsic shape is also reflected in the
distribution of $\lambda_{\mathrm{R}}$ and the ellipticities,
$\epsilon$, for random projections. In the left panel of
Fig. \ref{ell_lambda_ludwig_good} we show the 'observable'
projected ellipticities at the half-mass radius,  
$R_{1/2}$, for projections along the intermediate axis (edge-on,
filled circles), the long axis (open diamonds) and along the short
axis (face-on, open circles) versus the corresponding
$\lambda_{\mathrm{R}}$-parameter. All points are color coded according
to their edge-on rotation properties. Galaxies above an empirical
division line (dotted line in Fig. \ref{ell_lambda_ludwig_good},
\citealp{2011MNRAS.414..888E}) are fast rotators (blue) and those
below are slow rotators (red). The magenta solid line
\citep{2007MNRAS.379..418C} shows the edge-on view integrated up to
one effective radius for spheroids with an intrinsic anisotropy, $\beta
= 0.65 \times \epsilon$, which (including projection effects)
encompasses most observed fast rotating spheroids. The edge-on
projections cover 'observable' ellipticities in the range $0.16 <
\epsilon < 0.58$ and slow rotators on average have lower ellipticities
than fast rotators (this corresponds to the slightly higher values of
$c/a$ in Fig. \ref{shape_lambda_ludwig}). As soon as the galaxies are
projected along the long axis (blue open diamonds) the ellipticity and
$\lambda_{\mathrm{R}}$ of the fast rotators, which are typically close
to oblate, do not change much. In projection along the short axis
(face-on, blue open circles) most fast rotators appear round with very
low apparent angular momentum, as expected. For the slow rotators the
situation is slightly different. They are in general less axisymmetric
and seen along the long axis (red open diamonds) the ellipticity
already drops significantly to values around $\epsilon \sim 0.2$
whereas the low values for $\lambda_{\mathrm{R}}$ remain
unchanged. The apparent ellipticities are even more reduced when
'observed' along the short axis (open red circles). In some cases,
however, the ellipticity is even larger than in the projection along
the long axis (i.e. red open circles are at higher ellipticities than
red open diamonds).  

In the right panel of Fig. \ref{ell_lambda_ludwig_good} we again show
the location of the edge-on projection of the galaxies (solid circles)
in combination with 50 random projections (small dots). As expected
the projections populate the regions with lower ellipticities
and lower values of $\lambda_{\mathrm{R}}$. For the fast rotators
(blue) the projections follow the theoretical predictions for oblate
systems  with edge-on projections starting on the magenta line (black
dashed lines). The fact that  $\lambda_{\mathrm{R}}$ never drops below
a value of $\sim$ 0.05 is artificial and caused by the particle noise
in the simulations. For projections with low apparent (or real)
rotation adjacent Voronoi cells might have velocities that fluctuate
around zero. As $\lambda_{\mathrm{R}}$ is a cumulative parameter
of absolute values (bins with slightly negative as well as positive
velocities contribute) of all these velocities will be added up
creating a lower limit. If the galaxies are randomly projected 
(small dots) the fast rotators appear rounder with a mild decrease
in $\lambda_{\mathrm{R}}$ followed by a rapid drop when the ellipticities
approach zero. Projected slow rotators do not change their rotation
properties significantly but have lower apparent ellipticities. The
location of projected galaxies is in good agreement with predictions
from isolated binary merger simulations
\citep{2009MNRAS.397.1202J,2011MNRAS.416.1654B}.   

\begin{figure*}
\begin{center}
  \epsfig{file=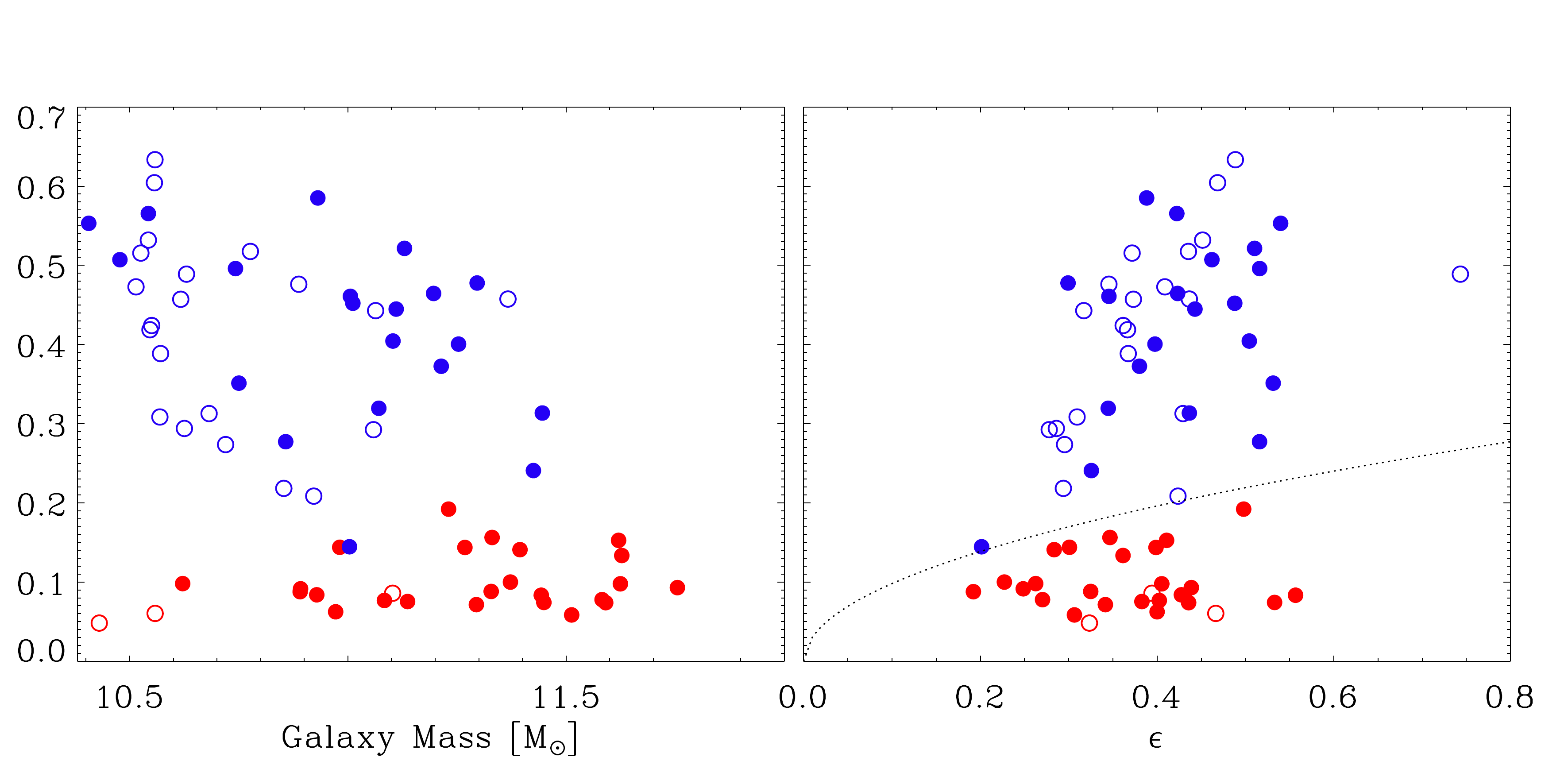,width=1.0\textwidth} 
  \caption{{\it Left:} The $\lambda_{\mathrm{R}}$ parameter measured
    in the edge-on projection for fast (blue) and slow rotators (red)
    versus the stellar mass of the simulated central galaxies 
    (filled circles) and satellite galaxies (open circles). Massive
    galaxies tend to be slow rotators. Less massive galaxies can be
    slow as well as fast rotators with fast rotators dominating the
    lower masses. Satellite galaxies (open circles) 
    follow a similar trend but are mainly fast rotators of lower
    mass. This is in qualitative agreement with $ATLAS^{\rm{3D}}$ galaxies
    \citep{2011MNRAS.414..888E}. {\it Right:} The
    $\lambda_{\mathrm{R}}$-parameter versus the projected
    ellipticities of the simulated central galaxies and 
    satellites (symbols as in the left figure). The black dotted division
    line between fast and slow rotators is based on the empirical
    relation presented in \citet{2011MNRAS.414..888E}. Satellite
    galaxies (open circles) follow the same trends as centrals. } 
\label{gm_lambdar_sat_ludwig}
\end{center}
\end{figure*}

Many of the central galaxies in our simulations have massive satellite
galaxies. We have identified all satellite galaxies with masses above
$10^{10} M_{\odot}$ and constructed their two-dimensional maps in the
same way as for the central galaxies. In Appendix \ref{satellitemaps}
we have collected the maps of all galaxies sorted by their host
galaxies IDs. In the left panel of
Fig. \ref{gm_lambdar_sat_ludwig} we show the distribution
of $\lambda_{\mathrm{R}}$ as a function of galaxy stellar mass for the 
central galaxies (solid circles) as well as the satellite galaxies
(open circles) color coded by their rotation properties. All satellite
galaxies have masses below $10^{10.8} M_{\odot}$ and most of them are
fast rotators and cover the same parameter space as the fast rotating
central galaxies (see also \citealp{2011MNRAS.417..845K}). Only three
of the satellite galaxies are slow rotators. On the right hand panel
we show the $\lambda_{\mathrm{R}}$-$\epsilon$ plane but now including the
satellite galaxies. In general, the trend for more massive galaxies
(now including the satellites) to rotate slower is in 
qualitative agreement with the observed $ATLAS^{\rm{3D}}$ galaxies (see
\citealp{2011MNRAS.414..888E}, Fig. 3). 

\section{Summary and Discussion}
\label{summary}
In this paper we present the first detailed two-dimensional
kinematic analysis of a sample of 44 cosmological zoom simulations of
massive galaxies. We investigate central galaxies in halos of 
2.2 $\times 10^{11} M_{\odot}$ - 3.7 $\times 10^{13} M_{\odot}$ with
stellar masses from 2.6 $\times 10^{10} M_{\odot}$ - 5.7 $\times
10^{11} M_{\odot}$ and their most massive satellites with a lower
stellar mass limit of $\times 10^{10} M_{\odot}$.  For every galaxy we
construct two-dimensional maps of the stellar line-of-sight velocity,
velocity  dispersion, and higher-order Gauss-Hermite moments $h_3$ and
$h_4$ \citep{1993ApJ...407..525V,1993MNRAS.265..213G} of the
line-of-sight velocity distribution. The analysis procedure is similar
to the one used for the integral field spectroscopic surveys $SAURON$
and $ATLAS^{\rm{3D}}$ \citep{2002MNRAS.329..513D,2011MNRAS.413..813C}.

The kinematic properties of the simulated galaxies - represented by
the two-dimensional velocity and dispersion fields - are in good 
qualitative agreement with the kinematics of observed nearby
elliptical galaxies in the $ATLAS^{\rm3D}$ survey. We find a similar
wealth of peculiar kinematic  features as for observed galaxies such as
counter-rotating cores (M0094),  slow rotation (M0125), kinematic
twists (M0215), minor axis rotation (M0190), dips in the central
velocity dispersion (M0224), misaligned rotation (M0224), dumbbell
features (M0858), enhanced dispersion along the major axis ($2\sigma$
feature, M0209), and regular disc-like rotation (M0408). The kinematic
maps for all galaxies and the most massive satellites can be found in
appendices \ref{samplemaps} and \ref{satellitemaps}.      

The global rotation properties are quantified with the
$\lambda_{\mathrm{R}}$-parameter. Some galaxies rotate slowly, 
some have significant rotation with disc-like velocity fields covering
almost the full range of the observed  $\lambda_{\mathrm{R}}$ values
within the effective (projected half-mass) radius of the galaxies
($0.05 < \lambda_{\mathrm{R}} <  0.6$), except for the fastest
rotators, which do not form in the simulations presented
here. However, the observed general trend for more massive galaxies to
rotate more slowly is apparent in the simulation sample. The
importance of this process for the present day early-type galaxy
population is not entirely clear. But empirical evidence indeed
suggests a continuity between the properties of spirals and fast
rotator ETGs on scaling relations \citep{2013MNRAS.432.1862C} 

A detailed analysis of the cosmological assembly (since $z \sim 2$) of
every individual galaxy enabled us to link the formation histories of
fast and slow rotators directly to specific features which show up in
the present day two-dimensional kinematic maps. The massive galaxies
in our sample can experience a larger number of mergers (up to
150 down to a mass-ratio of 1:100) and therefore a significant
fraction (30 to 50 percent)
of the stars finally residing in these galaxies has been accreted and
did not form inside the galaxies
\citep{2010ApJ...725.2312O,2012ApJ...744...63O,2012MNRAS.419.3200H}. We
find that the most prominent formation features characterising the final
rotation properties are the presence of late major mergers and the stellar
in-situ fraction, i.e. the ratio of stars that have formed in the galaxy (from
the accreted dissipative gas component) to the total amount of stellar
mass growth since $z \sim 2$  (other stars have been accreted in
mergers). All galaxies in our sample experience minor mergers, in
particular the massive ones, but major mergers (although rare) have (in
most cases) a  more dramatic effect on the rotation properties. The
effect of major mergers has been extensively investigated in the past
with idealised simulations and the results of these studies are
summarised in section \ref{theory}. 

In a cosmological context the formation histories of massive galaxies
are more complex and there are no two unique formation histories for
fast rotators and slow rotators, respectively. In fact we have to
consider several factors like merger mass-ratio, the timing of major
mergers, and gas fraction to uncover the different formation paths and
their present day kinematic imprints. We have identified
the following charactieristic formation histories resulting in the
formation of {\bf{fast rotators}}.

\begin{itemize}

\item Galaxies with gradual dissipation, a significant amount of
  stars formed in-situ (up to $\sim$ 40 per cent), and some late
  minor-mergers but no major mergers 
  since $z \sim 1$ (class A). All these galaxies show a peak in the
  $\lambda_{\mathrm{R}}$-profile inside one $r_e$  (a clear kinematic
  signature of dynamically cold discs), pointy isophotes, a
  depression of the stellar line-of-sight velocity dispersion along
  the major axis and characteristic dumbbell features. These fast
  rotators also have clearly asymmetric line-of-sight velocity
  profiles (steep leading wings) with anti-correlated  $v/\sigma$ and 
  $h_3$.   

\item Galaxies with a late gas-rich major merger leading to a net
  spin-up of the merger remnant or leaving a previously high-spin
  galaxy unchanged (class B). These galaxies have constantly rising
  $\lambda_{\mathrm{R}}$-profiles beyond $r_e$ most likely caused by
  the efficient angular momentum transfer during the major merger
  (e.g. \citealp{1992ARA&A..30..705B}). They 
  also have disc-like velocity fields, mid-plane depressions in
  the velocity dispersion, anti-correlated $h_3$ and $v/\sigma$, and,
  occasionally, pointy isophotes. This is generally consistent with 
  studies of isolated binary disc mergers (see section \ref{theory})
  indicating that under certain circumstances (e.g. proper orientation
  of progenitor discs  with high gas fractions, re-accretion of
  tidally stripped gas, strong feedback from star formation or gas
  accretion onto black holes, cooling from a hot, gaseous halo) even
  major merger remnants can have disc-like properties
  \citep{2002MNRAS.333..481B,2004ApJ...606...32R,2005ApJ...622L...9S,2006ApJ...645..986R,2011MNRAS.415.3750M}

\item Galaxies with one (or more) late ($z \lesssim 1$) dissipationless
  (stellar) major mergers (in-situ fraction $\lesssim 0.17$) leading to
  a significant spin-up of the stellar remnant or leaving the
  properties of a previously fast rotating galaxy unchanged (class D). These
  galaxies show moderate rotation ($\langle \lambda_{\mathrm{R}}
  \rangle \sim$ 0.23) but no additional signatures for embedded
  disc-like components and $v_{\mathrm{los}}$ and  $h_3$ are not
  anti-correlated. The absence of this anti-correlation is a generic 
  feature of rotating collisionless galaxy merger remnants
  \citep{2001ApJ...555L..91N,2006MNRAS.372..839N} and the potential
  spin-up in major galaxy mergers has been studied in detail with
  idealized experiments with qualitatively similar results
  \citep{2006ApJ...636L..81N,2009A&A...501L...9D,2010MNRAS.406.2405B,2011MNRAS.416.1654B}.   

\end{itemize}

We also identify three different formation paths for {\bf slow
  rotators}: 

\begin{itemize}
\item Galaxies with late ($z \lesssim 1$) gas-rich major mergers
  leading to a net spin-down of the progenitor galaxies (class C). These galaxies
  have high fractions of in-situ formed stars ($\sim$ 0.24), rotate slowly  ($\langle
  \lambda_{\mathrm{R}} \rangle \sim$ 0.1) and show characteristic
  central dips in the stellar velocity dispersion. In the idealized
  picture slowly rotating binary merger remnants of (gas-rich) disc
  galaxies have very similar properties to the galaxies in this class.
  Mergers like this have been traditionally considered (together with
  their collisionless counterparts, see below) as the origin of slow
  rotators 
  \citep{1996ApJ...471..115B,2006MNRAS.372..839N,2006ApJ...650..791C,2007MNRAS.376..997J,2009MNRAS.397.1202J,2009ApJ...705..920H,2010MNRAS.406.2405B,2010ApJ...723..818H,2011MNRAS.416.1654B}.   

\item   Galaxies that have undergone at least one recent recent gas
  poor (collisionless, dry) major merger (in addition to minor
  mergers), which has lead to a 
  significant spin-down of the remnant or has only mildly changed the
  properties of a previously slowly rotating galaxy (class E). Despite their slow
  rotation ($\langle \lambda_{\mathrm{R}} \rangle \sim$ 0.1, with
  slowly rising $\lambda_{\mathrm{R}}$ profiles at larger radii) the high
  ellipticities ($\langle \epsilon \rangle \sim$ 0.43) potentially
  cause tension with observations. Similar idealised mergers have also
  been considered to form slowly rotating spheroids
  \citep{1992ApJ...393..484B,1992ApJ...400..460H,2006ApJ...636L..81N} 
  but the potential tension with observed galaxies was realized
  early-on \citep{2003ApJ...597..893N,2006ApJ...650..791C}.

\item Galaxies for which the $z \lesssim 2$ mass assembly histories
  are dominated by stellar minor mergers without any major
  mergers (class F). These galaxies  are old ($\langle age \rangle \sim 10.9$
  Gyrs), have low in-situ fractions ($\sim 0.13$) and are the roundest
  ($\langle \epsilon \rangle \sim$ 0.27) and slowest ($\langle
  \lambda_{\mathrm{R}} \rangle \sim$ 0.08) rotators in our sample with 
  flat $\lambda_{\mathrm{R}}$ profiles and almost featureless velocity
  fields. They have continuously lost angular momentum since high
  redshift caused by repeated minor mergers, an interpretation that is
  supported by idealized minor merger simulations
  \citep{2007A&A...476.1179B,2010A&A...515A..11Q}. 

\end{itemize}

The above formation histories can be considered generic and are drawn
from simulations in a full cosmological context. They clearly provide more 
complex scenarios and a more complete picture than previous studies
based on idealized merger simulations. We have demonstrated that it is
possible to connect cosmological galaxy assembly histories to the
formation of fast and slow rotators with characteristic observable
higher-order kinematic features. Even though we consider the formation
paths as generic we cannot yet assess their full statistical
significance. The host halos (in which the simulated galaxies live)
were chosen to evenly cover an initially specified mass range and the
simulated sample is not large enough to represent the whole galaxy
population in this mass range. We address population properties in the
context of $ATLAS^{\rm{3D}}$ using semi-analytical models in
\citet{2011MNRAS.417..845K}. In particular, the sample is not large 
enough to finally interpret the observations of volume limited
surveys like $ATLAS^{\rm{3D}}$ or upcoming larger surveys and it is
beyond the scope of this paper to address galaxy population
properties. We rather investigate the possible variations in
individual formation histories, their impact on the kinematic features 
and global trends with mass. The results of the present study are 
in good agreement with a semi-analytical analysis presented in
\citet{2011MNRAS.417..845K}. In particular the role of dissipation and
the complex formation histories proposed in the semi-anlaytical
analysis are confirmed by the present study. 

In addition to the successes of the models presented here we would
like to draw the attention to the model limitations. The simulated
halo sample was randomly selected from a larger cosmological box and
is inherently different to the volume limited sample of observed
galaxies. This is a potential source of systematic error. All cosmological
simulations of galaxy formation either only include a subset of the
relevant physical processes or make simplified assumptions about the
unresolved physics of star formation, black hole formation and the
respective feedback, background radiation, cooling etc. Therefore the
emerging results are inevitably, to some degree, model dependent and
require careful interpretation. We would like to use this discussion
to give the necessary critical review of our results.   

As detailed in Section \ref{simulations} our simulations use an algorithm for
star formation which, by construction, favors the early formation of
stars in small subunits and we form no bona-fide late type galaxies in
our sample. Feedback from supernovae is included, but
does not drive strong winds. Other studies with similar simulation
parameters typically find simulated galaxies with photometric and
kinematic properties similar to present day early-type galaxies
\citep{2007ApJ...658..710N,2009ApJ...697L..38J,2009ApJ...699L.178N,2010ApJ...709..218F,2012arXiv1202.3441J}.
The properties of the stellar components of the galaxies presented
here agree with present day early-type galaxy scaling relations of mass with radius
and stellar velocity dispersion \citet{2012ApJ...744...63O}. In addition, they
have close to isothermal total mass distributions
\citep{2012MNRAS.423.1813L,2013ApJ...766...71R}, similar to observed
early-type galaxies  (e.g. \citealp{2011MNRAS.415.2215B}), and their
observed size evolution between $z \sim 2$ and $z = 0$ is in agreement
with recent observational estimates
\citep{2012ApJ...744...63O,2012MNRAS.423.1813L}. However,
as the employed star formation model favors efficient star formation
at high redshift we preferentially form spheroidal systems with
predominantly old stellar populations. There is growing theoretical
and observational evidence that the suppression of early star
formation by strong feedback processes (which are not included in the
present study) leads to more realistic formation
paths for disc-like galaxies at low as well as high redshift
(e.g. \citealp{2012ApJ...745...11G,2012arXiv1211.1021H,2012arXiv1211.3120H,2013arXiv1303.6959A,2013MNRAS.428..129S, 
2013arXiv1302.2618K,2013arXiv1304.1559A}). The 
importance of this process for the present day early-type galaxy
population is not entirely clear. Feedback can promote the formation
of a population of highly flattened galaxies with very fast rotation
\citet{2013arXiv1304.1559A,2013arXiv1305.5360M}. However, for these
galaxies to become early-type they have to stop star formation. How
this might work is poorly understood and can depend on environment and
mass but also on details of the model implementation for supernova and
AGN feedback. The galaxies studied here have on average old stellar
populations (see Tab. \ref{tab:classes}) and late star formation is not a
major problem. A good way of testing this particular aspect of the
simulations (star formation, gas, quenching) is to compare the gas
properties of simulated galaxies at z=0 with the observed properties
(including HI, CO, ionised gas, and X-ray gas).

In addition, the fraction of available baryons (in every halo)
converted into stars in the central galaxies for simulated massive
galaxies presented here is typically 2 to 5 times higher than
estimates from models matching observed galaxy mass  functions to
simulated halo mass functions
\citep{2010ApJ...725.2312O,2010MNRAS.404.1111G,2010ApJ...710..903M,2010ApJ...717..379B,2011MNRAS.416.1486N,2011ApJ...738...45L,2012ApJ...752...41Y,2012arXiv1207.6105B,2013MNRAS.428.3121M}. In
addition, we form no bone-fide late type galaxies in our
sample. However, due to the random selection of halos some fraction of
the of the lower mass galaxies in our sample should be star forming
disc like systems. Possible physical processes responsible for this discrepancy, and missing in
our models, are again feedback from SNII driving galactic
\citep{1974MNRAS.169..229L,1986ApJ...303...39D,2008MNRAS.389.1137S,2008MNRAS.387..577O,2010Natur.463..203G,2010MNRAS.402.1599S,2012arXiv1203.5667D}
and/or feedback from super-massive black holes
\citep{2006MNRAS.365...11C,2008ApJ...676...33D,2009MNRAS.400..100S,2010MNRAS.406..936P,2010ApJ...722..642O,2010MNRAS.406..822M,2011MNRAS.414..195T,2013MNRAS.428.2966P,2013arXiv1305.2913V}. Many
simulations have demonstrated that feedback processes can expel the
baryons from galaxies and influence the ratio of in-situ formed to
accreted stars and the morphology of the systems 
\citep{2012MNRAS.419.3200H,2012MNRAS.425..641L,2012MNRAS.423.1544S,2013arXiv1301.3092D}. However,
only a few  attempts have been made to study these processes in more
detail for massive galaxies
\citep{2011MNRAS.414..195T,2013arXiv1301.3092D,2012MNRAS.422.3081M,2013MNRAS.428.2966P,2013arXiv1305.2913V}. For
example \citet{2012MNRAS.422.3081M,2012MNRAS.420.2859M,2013arXiv1301.3092D}  
demonstrate that AGN feedback can impact the kinematic properties the
stars in the central parts of massive galaxies and promote the
formation of slow rotators.  

Despite the modelling uncertainties we consider the direct connection
between cosmological assembly histories of massive galaxies and their
present day two-dimensional kinematic features presented in this paper
to be generally valid.

\section*{Acknowledgments}
We thank the anonymous referee for valuable comments on the
manuscript. MC acknowledges support from a Royal Society University Research Fellowship. This work was
supported by the rolling grants `Astrophysics at Oxford' PP/E001114/1
and ST/H002456/1 and visitors grants PPA/V/S/2002/00553, PP/E001564/1
and ST/H504862/1 from the UK Research Councils. RLD acknowledges
travel and computer grants from Christ Church, Oxford and support from
the Royal Society in the form of a Wolfson Merit Award
502011.K502/jd. LD is also grateful for support from the Australian
Astronomical Observatory Distinguished Visitors programme, the ARC
Centre of Excellence for All Sky Astrophysics, and the University of
Sydney during a sabbatical visit. SK acknowledges support 
from the the Royal Society Joint Projects Grant JP0869822. RMcD is
supported by the GeminiObservatory, which is operated by the
Association of Universities for Research in Astronomy, Inc., on behalf
of the international Gemini partnership of Argentina, Australia,
Brazil, Canada, Chile, the United Kingdom, and the United States of
America. MB has received, during this research, funding from the
European Research Council under the Advanced Grant Program Num
267399-Momentum. PS is a NWO/Veni fellow. MS acknowledges 
support from a STFC Advanced Fellowship ST/F009186/1. LY acknowledges
support from NSF grant AST-1109803. The authors acknowledge 
financial support from ESO. 
\clearpage 

\bibliographystyle{mn2e}
\bibliography{../../REFERENCES/references}

\appendix 
\section{Kinematic maps of the central galaxies} 
\label{samplemaps} 

In this section we show all two-dimensional kinematic maps of the
central galaxies in the simulated halos. The galaxies can be
identified by their number and the global galaxy properties are given
in Tab. \ref{tab1}. The maps are constructed as described in section
\ref{maps} and we present the line-of-sight velocity (left panels),
velocity dispersion (second to left), $h_3$ (second to right), and
$h_4$ (right panels). 

\begin{figure*}
\begin{center}
  \epsfig{file=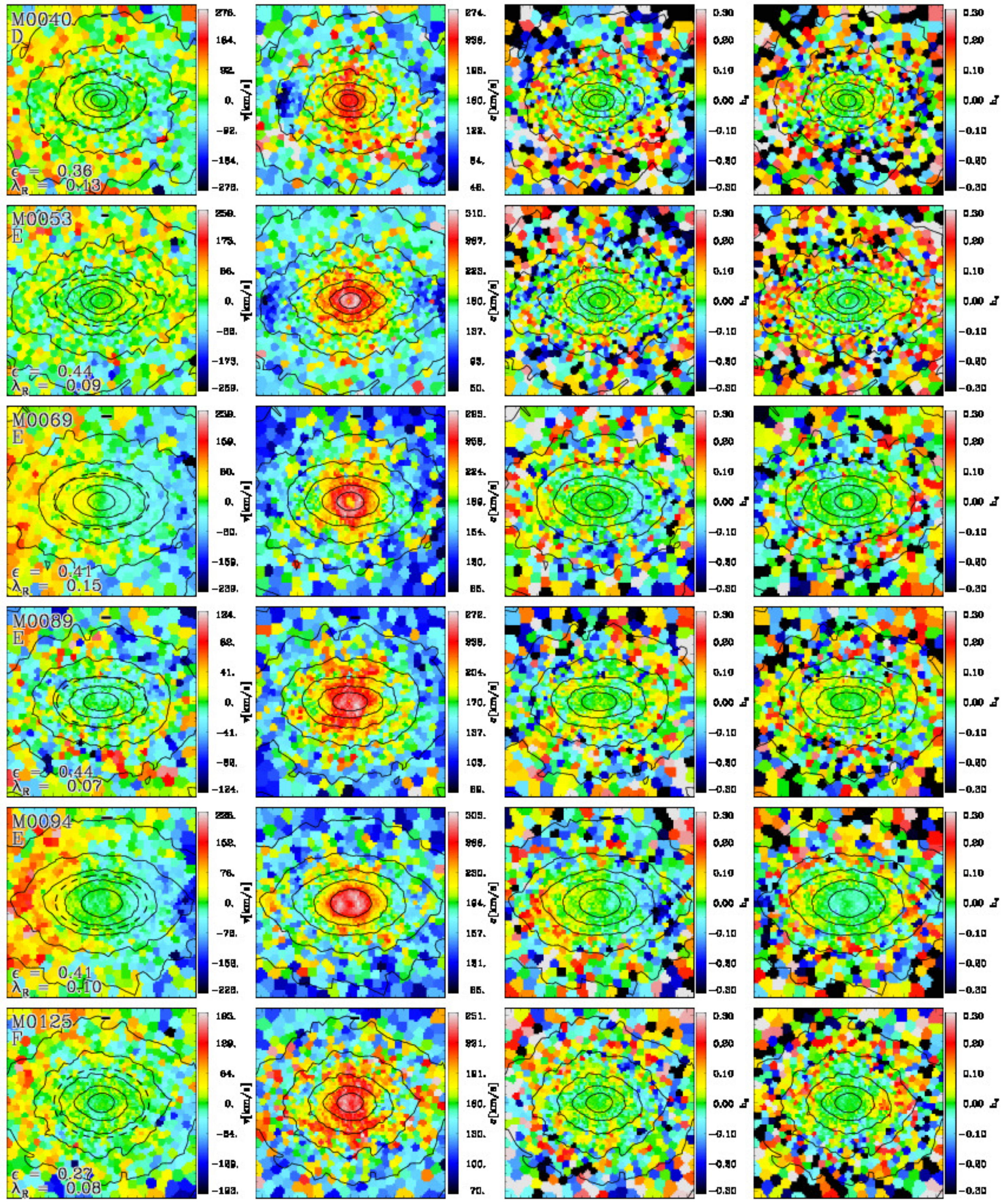, width=\textwidth}
  \caption{Two-dimensional kinematic maps of the simulated
    galaxies. From left to right we show the line-of-sight velocity,
    velocity dispersion, $h_3$, $h_4$. The central galaxies are named
    by their identification number (e.g. M0040). Isophotal contours
    are indicated by the black lines and the projected ellipticity at
    $r_{\mathrm{e}}$ as well as $\lambda_{\mathrm{R}}$ is given in the
    left panels (see also Tab. \ref{tab1} for galaxy properties). The
    bar indicates 1kpc.   
}
\label{maps0}
\end{center}
\end{figure*}

\begin{figure*}
\begin{center}
  \epsfig{file=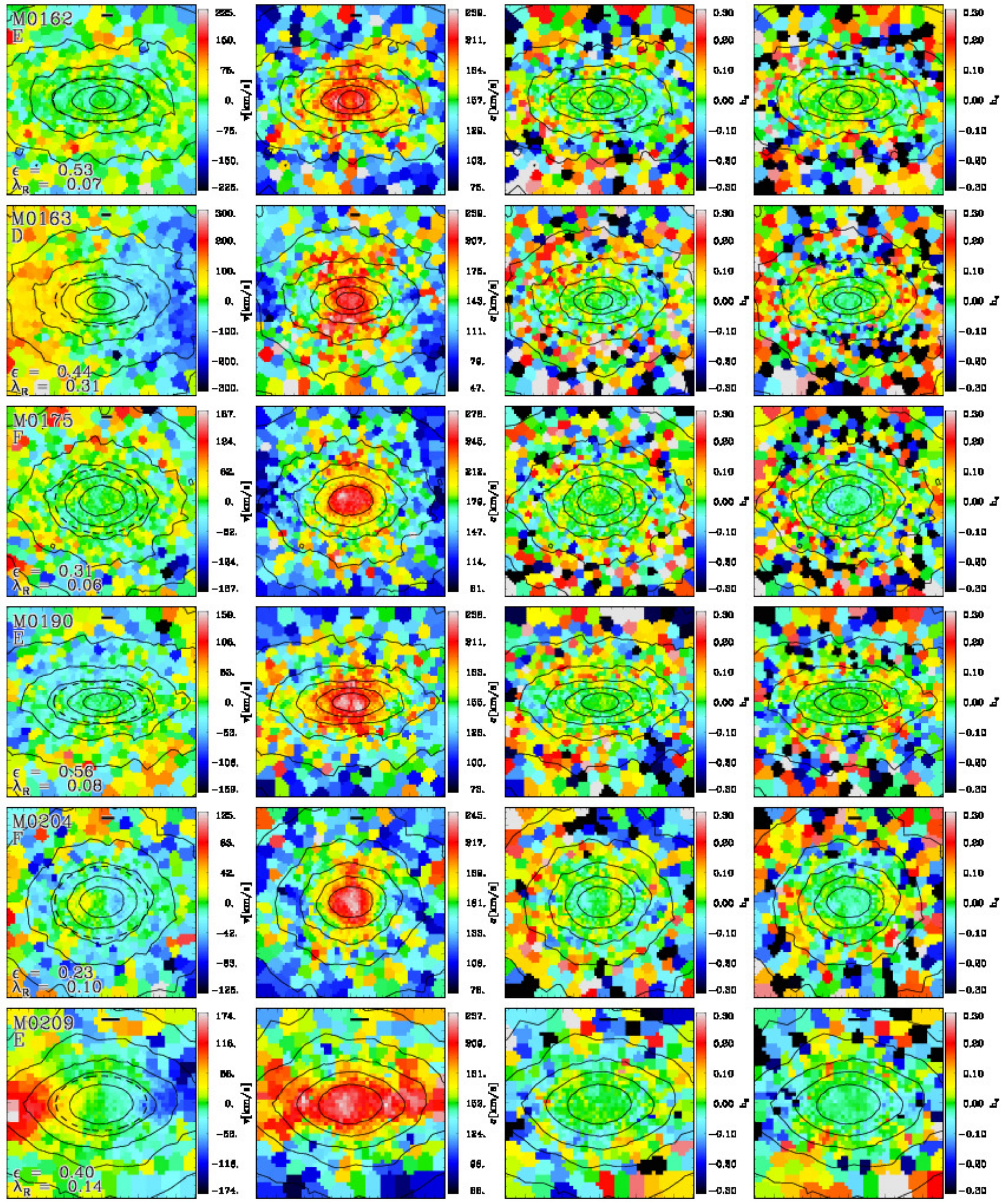, width=\textwidth}
  \caption{Same as Fig. \ref{maps0} for galaxies M0162 to M0209}
\label{maps1}
\end{center}
\end{figure*}

\begin{figure*}
\begin{center}
  \epsfig{file=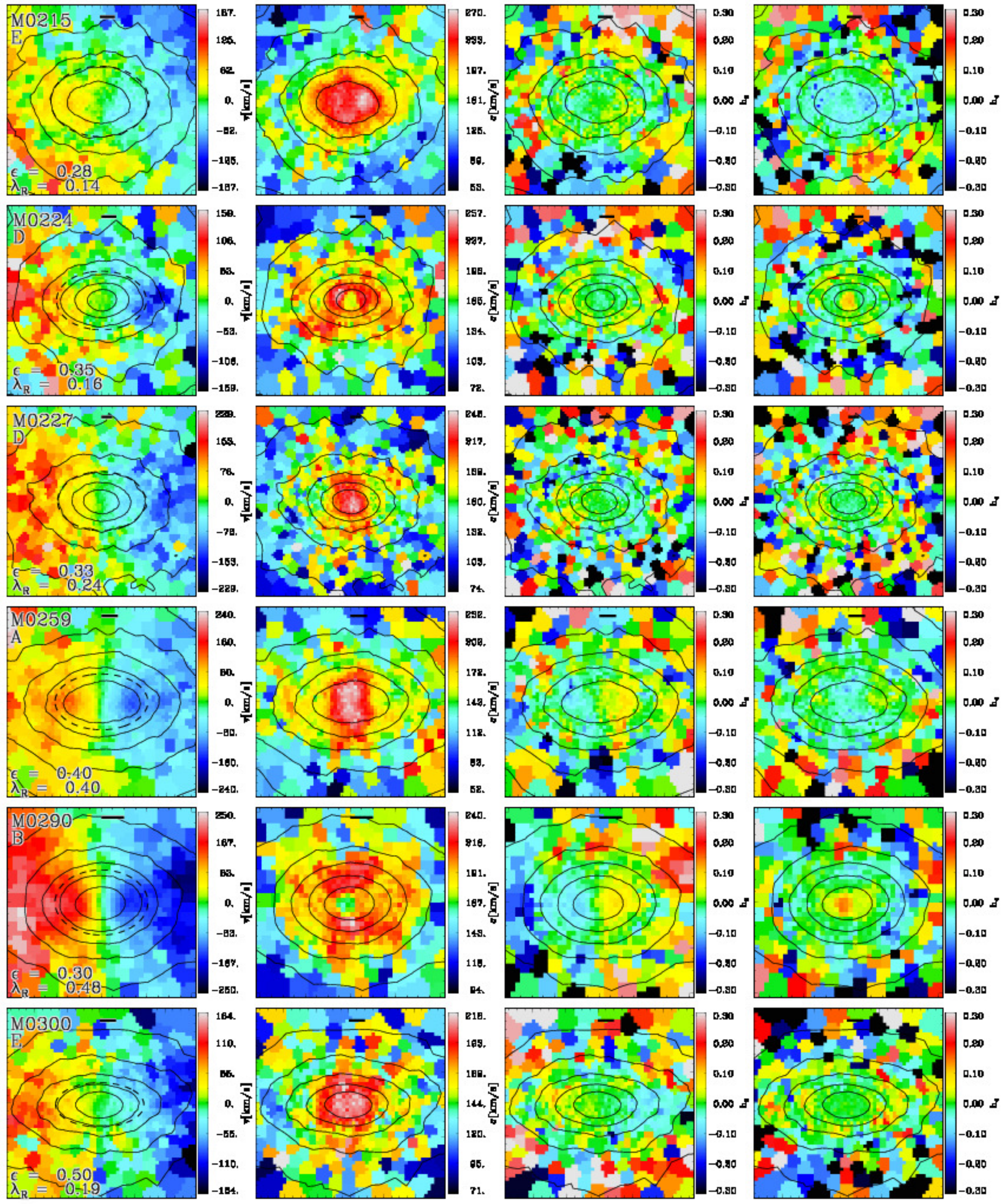, width=\textwidth}
  \caption{Same as Fig. \ref{maps0} for galaxies M0215 to M0300}
\label{maps2}
\end{center}
\end{figure*}

\begin{figure*}
\begin{center}
  \epsfig{file=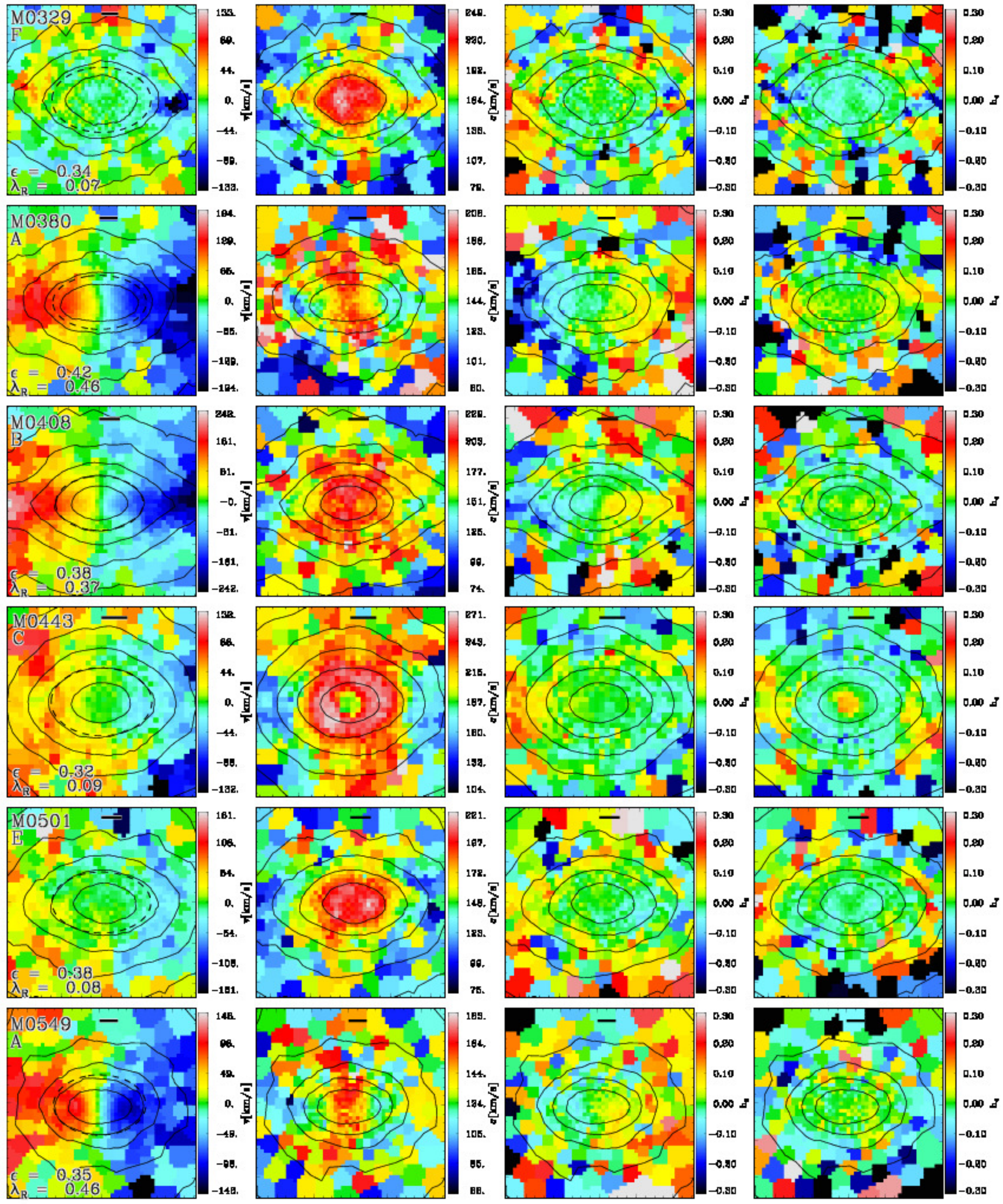, width=\textwidth}
  \caption{Same as Fig. \ref{maps0} for galaxies M0329 to M0549}
\label{maps3}
\end{center}
\end{figure*}

\begin{figure*}
\begin{center}
  \epsfig{file=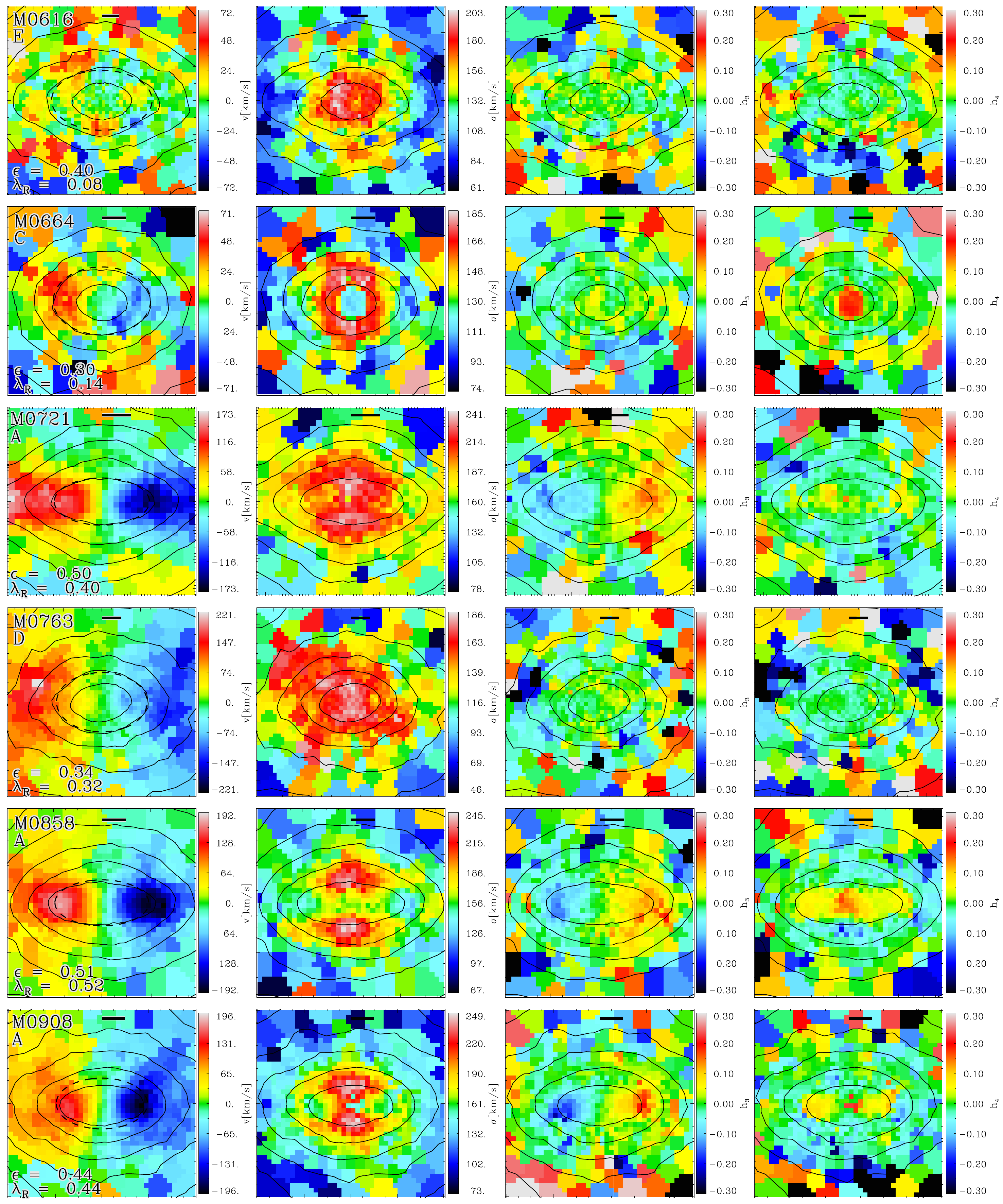, width=\textwidth}
  \caption{Same as Fig. \ref{maps0} for galaxies M0616 to M0908}
\label{maps4}
\end{center}
\end{figure*}

\begin{figure*}
\begin{center}
  \epsfig{file=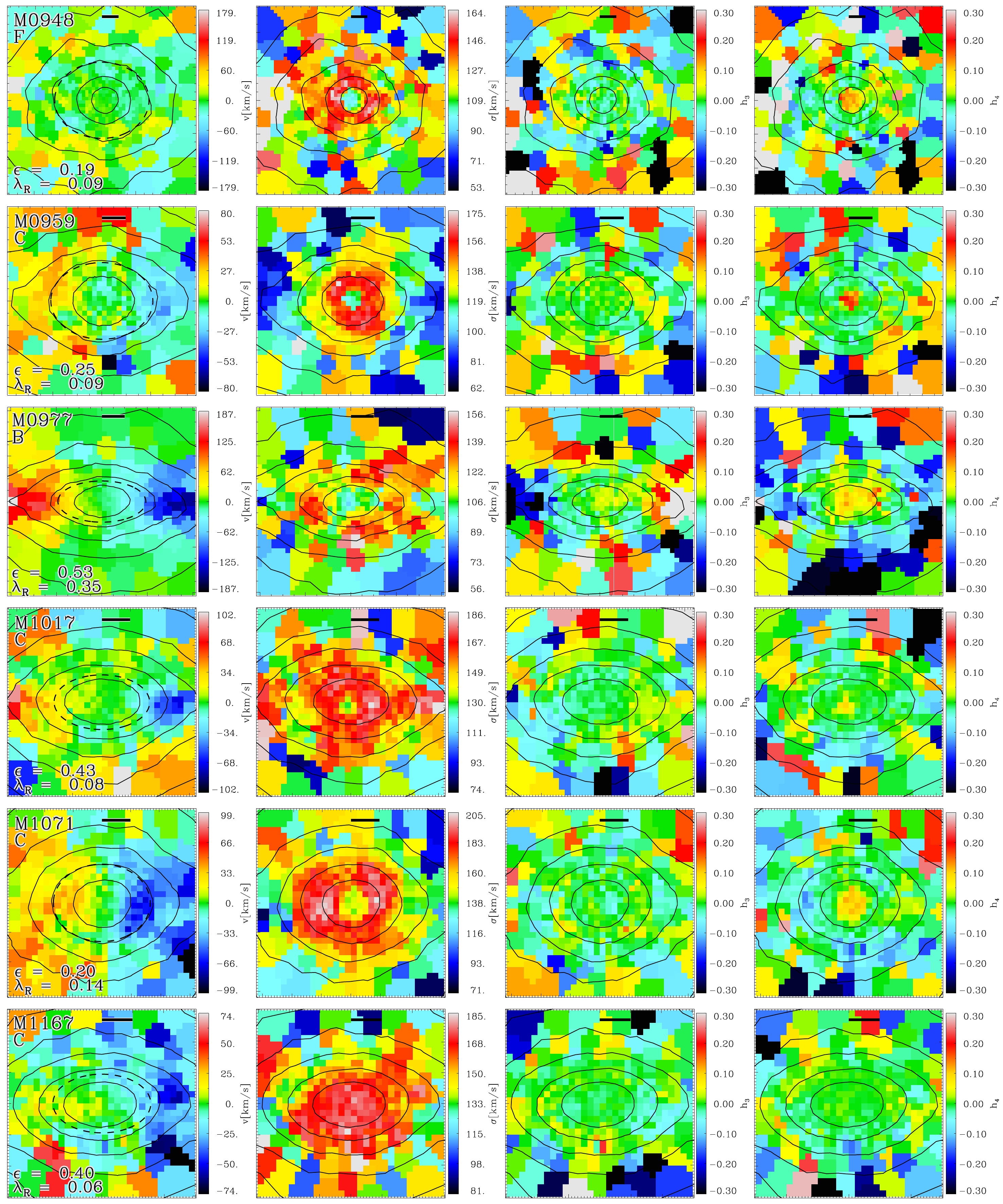, width=\textwidth}
  \caption{Same as Fig. \ref{maps0} for galaxies M0948 to M1167}
\label{maps5}
\end{center}
\end{figure*}

\begin{figure*}
\begin{center}
  \epsfig{file=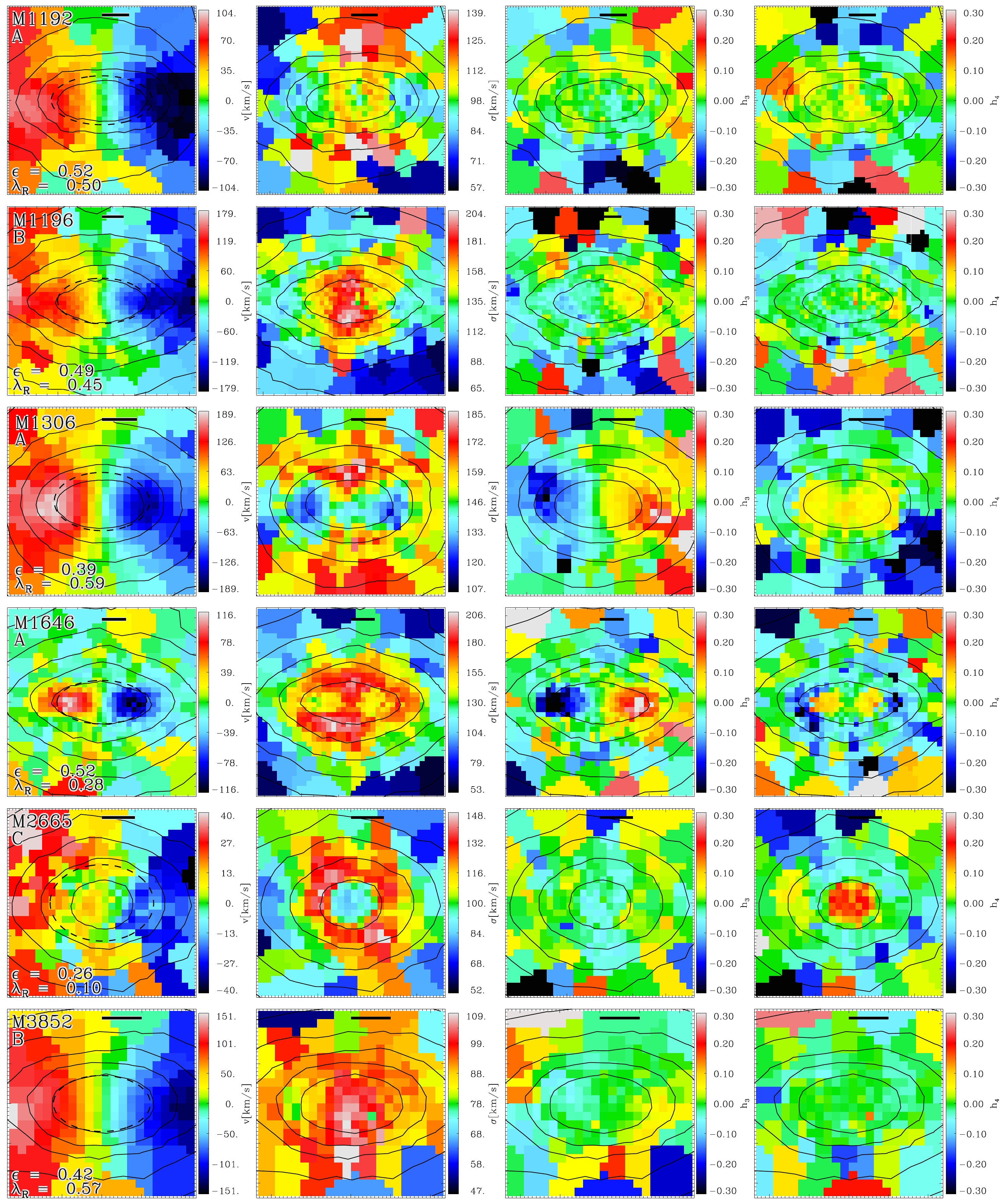, width=\textwidth}
  \caption{Same as Fig. \ref{maps0} for galaxies M1192 to M3852}
\label{maps6}
\end{center}
\end{figure*}

\begin{figure*}
\begin{center}
  \epsfig{file=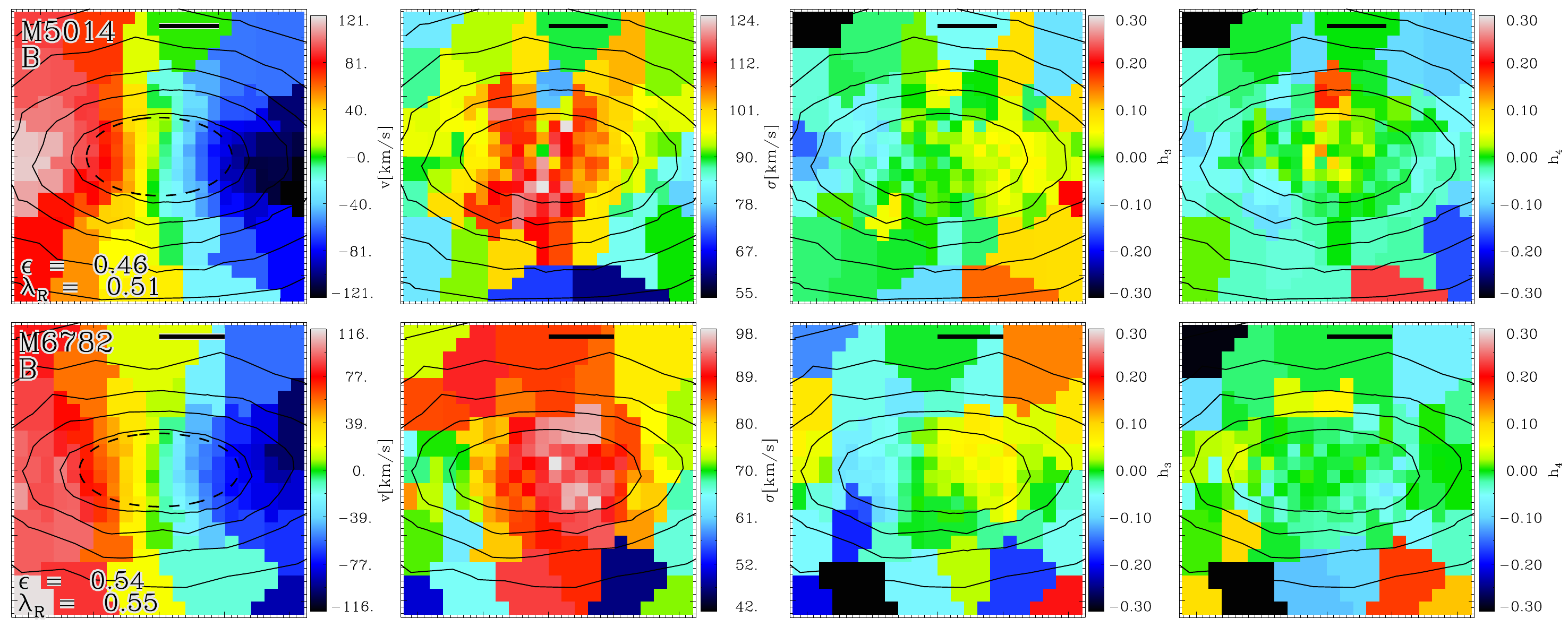, width=\textwidth}
  \caption{Same as Fig. \ref{maps0} for galaxies M5014 and M6782}
\label{maps7}
\end{center}
\end{figure*}

\clearpage 

\section{Kinematic maps of the satellite galaxies} 
\label{satellitemaps} 

Here we show all two-dimensional kinematic maps of the most massive 
satellites down to a stellar mass limit of ???. The host galaxy of the
satellites is indicated by the main galaxy number followed by -S1,
-S2 etc., e.g. M0040-S3 is the third most massive satellite in the
dark matter halo of galaxy M0040. The maps are constructed as
described in section \ref{maps} and we present the line-of-sight
velocity (left panels), velocity dispersion (second to left), $h_3$
(second to right), and $h_4$ (right panels). 

\begin{figure*}
\begin{center}
  \epsfig{file=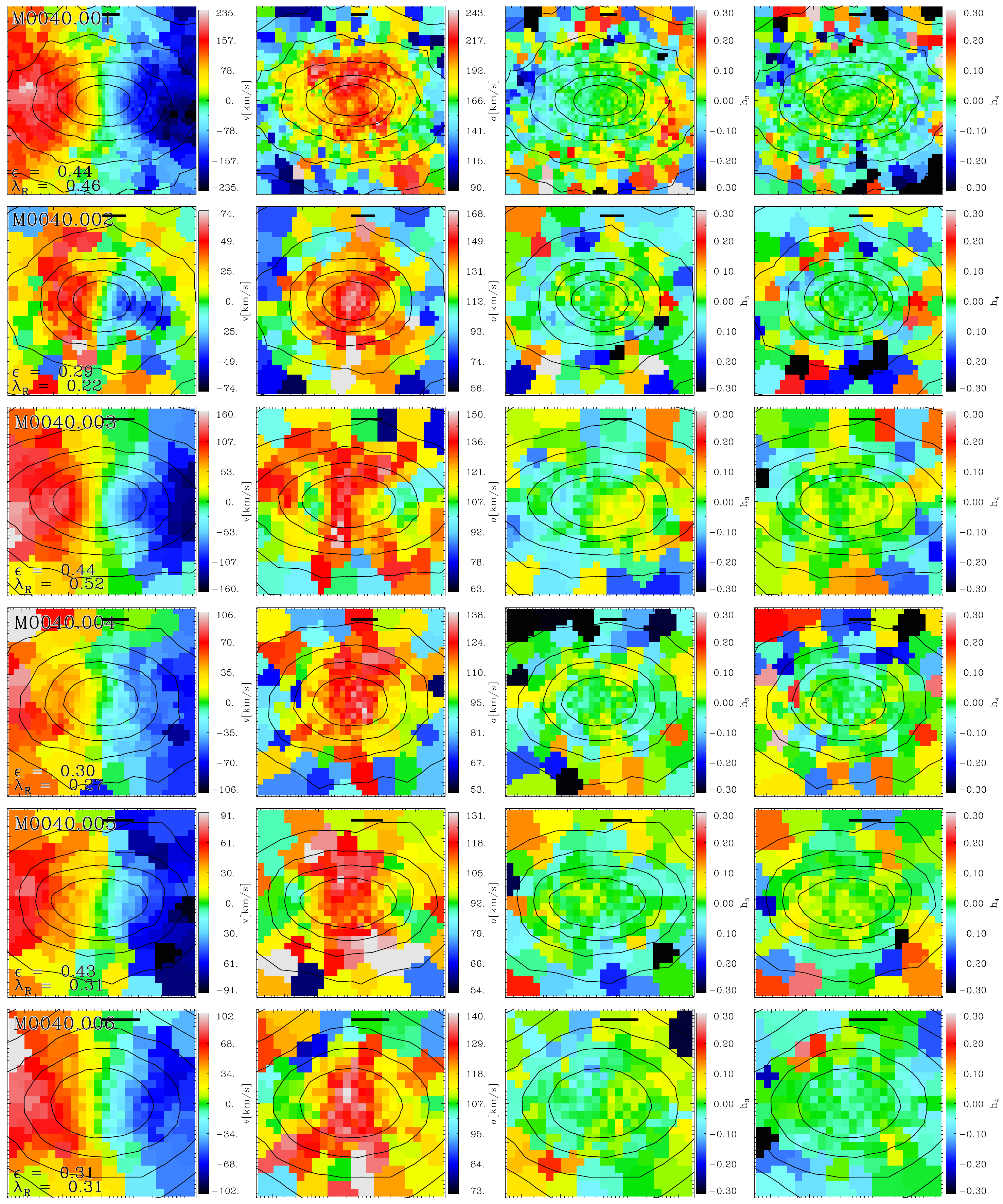, width=\textwidth}
  \caption{Two-dimensional kinematic maps of the simulated
    most massive satellite galaxies. From left to right we show the
    line-of-sight velocity, velocity dispersion, $h_3$, $h_4$. The
    satellite galaxies residing in the main galaxy halo are named
    according to the identification number of the main galaxy
    (e.g. M0040) followed by an  identification number ordered by
    stellar mass and indicated by a three digit number .001, .002
    etc. E.g. M0040.002 is the second most 
    massive satellite galaxy in the dark matter halo of
    M0040. Isophotal contours are indicated by the black lines and the
    projected ellipticity at $r_{\mathrm{e}}$ as well as
    $\lambda_{\mathrm{R}}$ is given in the left columns. The bar
    indicates a physical size of 1kpc.  
}
\label{satmaps0}
\end{center}
\end{figure*}

\begin{figure*}
\begin{center}
  \epsfig{file=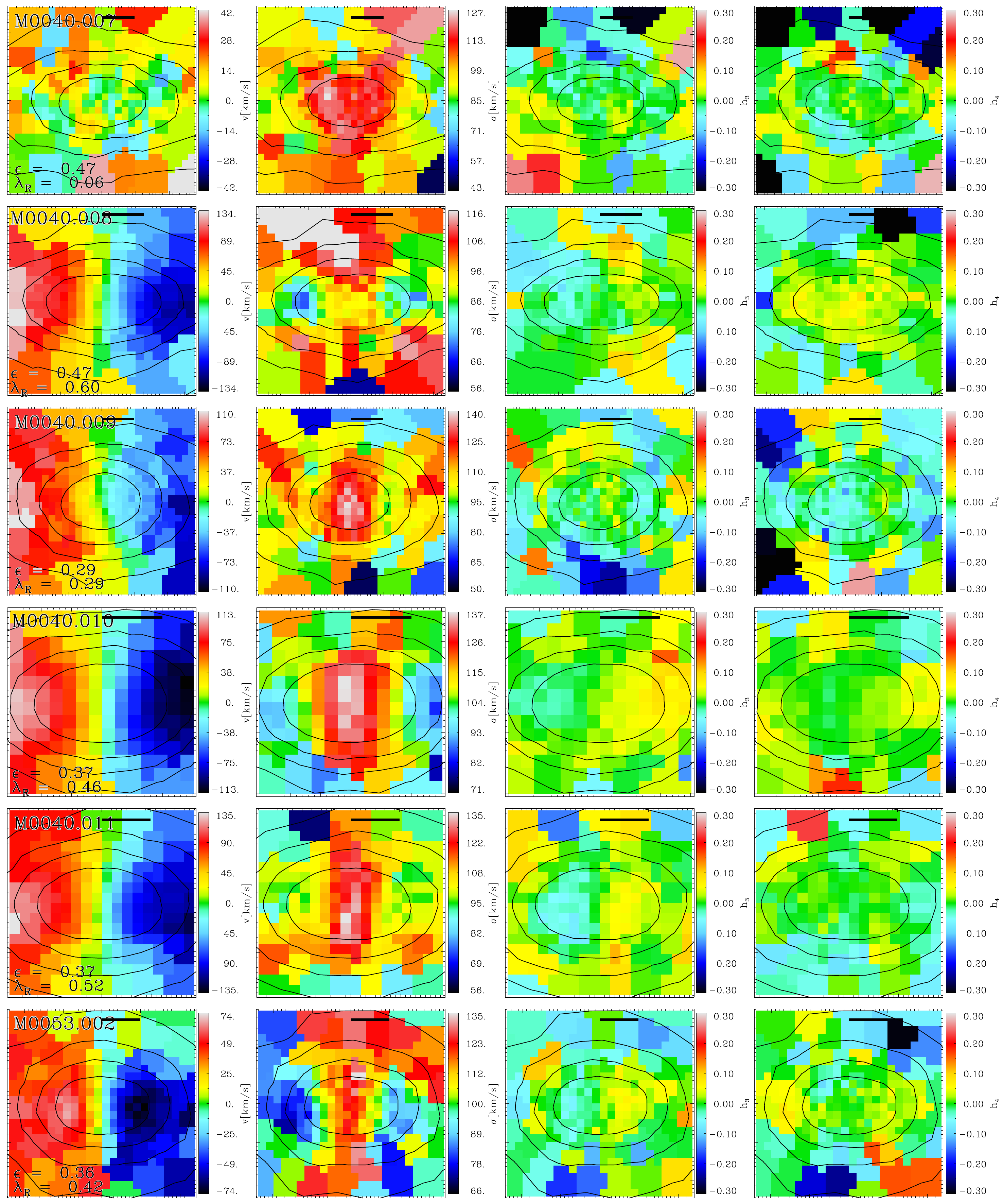, width=\textwidth}
  \caption{Same as Fig. \ref{maps0} for the satellite galaxies of
    M0040 and M0053.}
\label{satmaps1}
\end{center}
\end{figure*}

\begin{figure*}
\begin{center}
  \epsfig{file=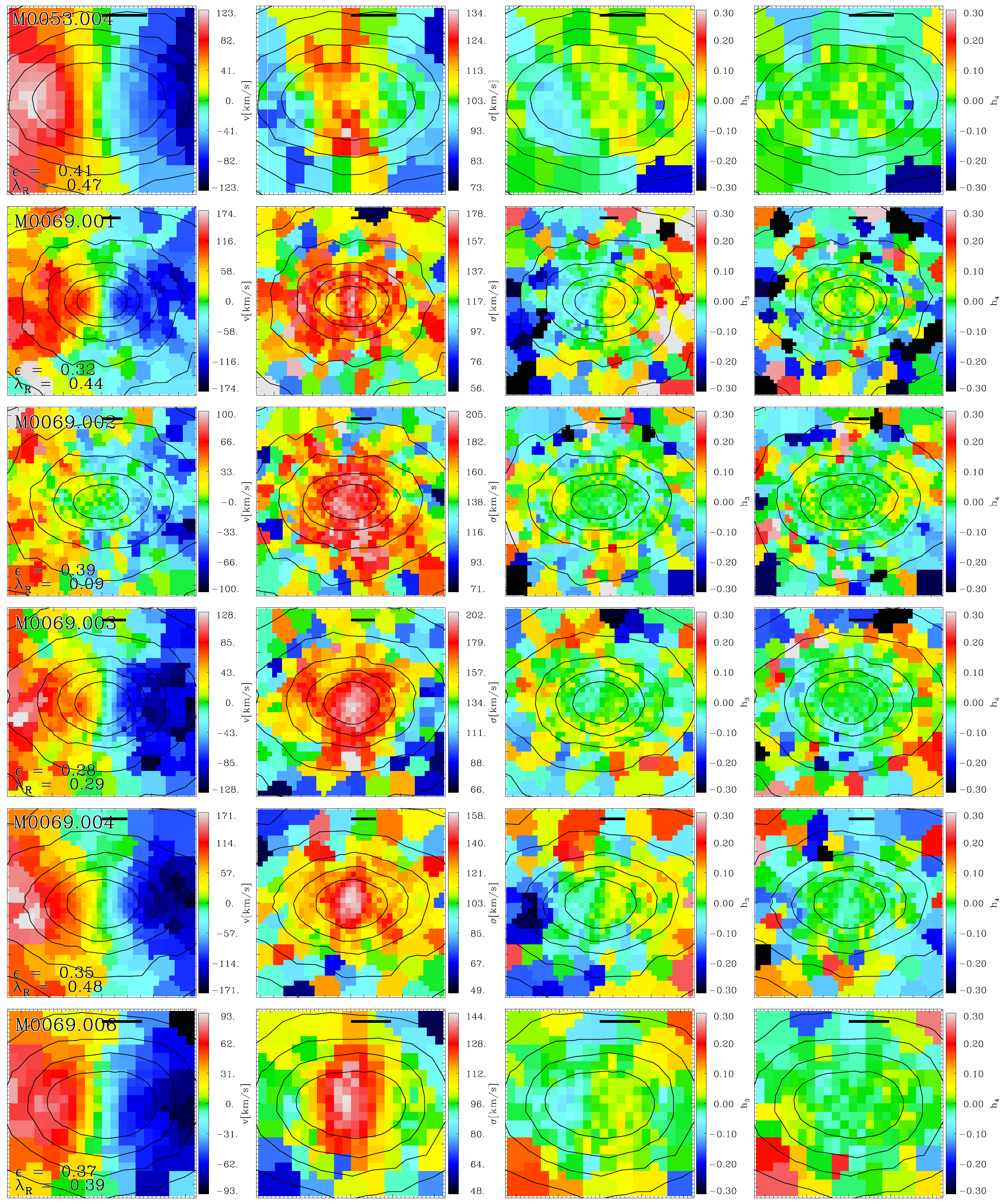, width=\textwidth}
  \caption{Same as Fig. \ref{maps0} for the satellite galaxies of
    M0053 and M0069.}
\label{satmaps2}
\end{center}
\end{figure*}

\begin{figure*}
\begin{center}
  \epsfig{file=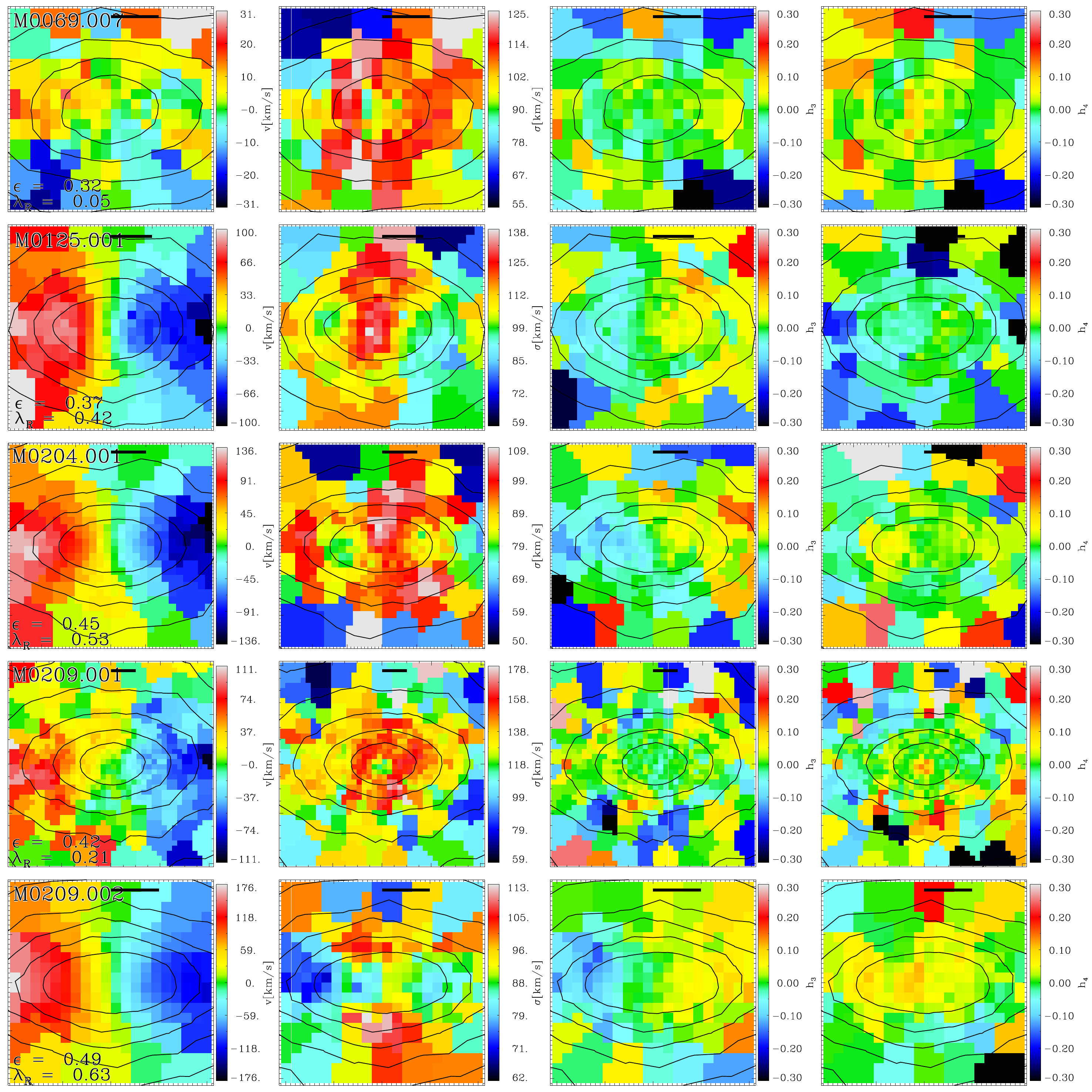, width=\textwidth}
  \caption{Same as Fig. \ref{maps0} for galaxies of M0069, M0125,
    M0204, and M0209.}
\label{satmaps3}
\end{center}
\end{figure*}

\label{lastpage}
\end{document}